\documentclass{aastex63}

\usepackage{epsfig}
\usepackage{epsfig,colortbl,natbib}
\usepackage{amsfonts}
\usepackage{amssymb}
\usepackage{amsthm}
\usepackage{amsmath}
\usepackage{times}
\usepackage{wrapfig}
\usepackage{color}
\usepackage{soul}
\usepackage{multirow}
\usepackage[utf8]{inputenc}

\received{...}
\revised{...}
\accepted{...}
\submitjournal{ApJ}

\shorttitle{}
\shortauthors{}
\graphicspath{{./}{figures/}}

\begin{document}

\title{Effect of acceleration and escape of energetic particles on spectral steepening at shocks}

\correspondingauthor{Federico Fraschetti}
\email{ffrasche@lpl.arizona.edu}

\author[0000-0002-5456-4771]{Federico Fraschetti}
\affiliation{Department of Planetary Sciences-Lunar and Planetary Laboratory, University of Arizona, Tucson, AZ, 85721, USA }
\affiliation{Center for Astrophysics $|$ Harvard \& Smithsonian, Cambridge, MA, 02138, USA}

\begin{abstract}
Energetic particles spectra at interplanetary shocks often exhibit a power law within a narrow momentum range softening at higher energy. We introduce a transport equation accounting for particle acceleration and escape with diffusion contributed by self-generated turbulence close to the shock and by pre-existing turbulence far upstream. 
The upstream particle intensity steepens within one diffusion length from the shock as compared with diffusive shock acceleration rollover. The momentum spectrum, controlled by macroscopic parameters such as shock compression, speed, far upstream diffusion coefficient and escape time at the shock, can be reduced to a log-parabola and also to a broken power law. In the case of upstream uniform diffusion coefficient, the largely used power law/exponential cut off solution is retrieved. 
\end{abstract}

\section{Introduction}

The diffusion of charged particles at shock waves has been regarded for decades as the major process forging the acceleration of energetic particles \citep[e.g.,][]{Drury:83}, with a resulting power-law momentum spectrum  \citep{Axford.etal:77,Bell:78a,Blandford.Ostriker:78,Krymskii:77}. A commonplace steepening from single power-law over at least 2 decades in particle momentum is routinely measured {\it in-situ} by monitoring suprathermal particles during the arrival of interplanetary shocks to detectors at multiple locations at $1$ AU \citep{Lario.etal:19}, or in the time-integrated fluence spectra of energetic particles in Ground Level Enhancements \citep[GLEs,][]{Mewaldt.etal:12} or in astrophysical observations of radio and $X$-/$\gamma$-rays emitted respectively by GeV and TeV electrons in supernova remnants \citep{Li.etal:18} or pulsar wind nebulae \citep{Meyer.etal:10}. The question as to what mechanisms govern the unequivocally observed bending spectrum, e.g., escape, transport effects, spherical expansion, shock deceleration and their relative contribution in individual observations remains unanswered; it is also crucial to determine the relative importance of a diffusive shock acceleration (DSA) version that includes escape as compared to, e.g., downstream magnetic reconnection \citep{Zank.etal:14}.

The widespread power law + exponential cut-off spectral fitting model \citep{Ellison.Ramaty:85} encapsulates diffusive confinement into the spectral energy break. Double power-laws in solar energetic particle (SEP) spectra have been used to fit time-integrated spectra for the past two decades \citep[e.g.][]{Mason.etal:02, Mewaldt.etal:05b, Desai.etal:16}. Multi-spacecraft {\it in-situ} spectra for GLEs are satisfatorily fit \citep{Mewaldt.etal:12} by the empirical broken-power-law model introduced by \cite{Band.etal:93} for the astrophysical Gamma-Ray Bursts spectra \citep[e.g.,][]{Fraschetti.etal:06}. \cite{Cohen.Mewaldt:18} found that the 2017 September $10^{th}$ GLE event spectrum is well fitted by a broken power law. The effect of a finite spatial region in the low corona on the particle acceleration at shocks driven by Coronal Mass Ejections (CME) was investigated \citep{Schwadron.etal:15} by solving a transport equation that accounts for catastrophic losses, i.e., particle escape from the region where DSA applies, and a broken power-law solution was found.  \citet{Li.Lee:15} have shown that the Parker equation with power-law injected near the Sun admits a double-power-law as a solution at a distance of $1$ AU if an energy dependence of the scattering mean free path and adiabatic deceleration in the radially divergent solar wind (SW) are included; however, such study neglects the perpendicular diffusion, that was shown to contribute to the longitudinal spread in large gradual SEP events \citep[and references therein]{Dresing.etal:12,Droege.etal:16} and to drifts \citep{Fraschetti.Jokipii:11,Dalla.etal:13}. From  three {\it IMP8} (Interplanetary Monitory Platform 8) events occurred on September $19^{th}$, November $22^{nd}$ 1977 and March $1^{st}$ 1979, \cite{Zhao.etal:17}  assessed that the transport across the interplanetary turbulence plays a relevant role, in comparison with the acceleration, in the break of the energetic particles spectra. \cite{Bruno.etal:18} have shown that the $\sim 80$ MeV-few GeV spectra of ions collected by {\it PAMELA} (Payload for Antimatter Matter Exploration and Light-nuclei Astrophysics) for SEP events are satisfactorily fitted with a power law + exponential rollover. An escape-free model by \cite{Drury:11} yields a spectral break from the combined effect of 1D radial shock expansion and decrease of the acceleration time scale by parametrizing particle injection into DSA. The semi-analytic solution of a model for acceleration and trapping of energetic ions at parallel shocks propagating between $5.8$ and $60$ solar radii is well fitted with a power law + exponential cutoff \citep{Vainio.etal:14}. Time-integrated broken power law spectra, by implementing the nested shell blast wave approach  \citep{Zank.etal:00a,Li.etal:03,Rice.etal:03}, are supported by {\it ACE} and {\it Ulysses} data  \citep{Verkhoglyadova.etal:09}. Recently,  \cite{Malkov.Aharonian:19} presented a time-dependent model for the escape-free energetic particles intensity across shocks in 3D large-scale spherical geometry expanding in a homogeneous magnetic field and analytically found a spectral steepening. Penetration of astrophysical shocks into neutral-rich molecular clouds was found to steepen the spectrum due to ion-neutral wave-damping \citep{Malkov.etal:11}. Spectra of spike events at interplanetary shocks \citep{Lario.etal:03} are found numerically to deviate from power law \citep{Fraschetti.Giacalone:15} and can be fitted with a particular form of non-power law spectrum, i.e., Weibull function, assuming a time-dependent leaky-box model with a second-order Fermi acceleration  \citep{Pallocchia.etal:17}.

In the past two decades, the spectral model of log-parabola was shown to reproduce a broad variety of observations: protons spectra of the 16 GLE events of solar cycle 24 are well fitted by a log-parabola  \citep{Zhou.etal:18}. In the astrophysical context, photon spectra in the  $X-$ray band of blazars \citep{Massaro.etal:04} are best-fitted by assuming a log-parabola spectrum of synchrotron-emitting electrons; a significant fraction of photon spectra of extended sources (e.g., supernova remnants) in the third {\it Fermi}/Large Area Telescope (LAT) source catalog \citep{Fermicoll:15a} are best-fitted by a log-parabola rather than a single power-law. The photon spectrum of the Crab Nebula in the range $(10^{-5} - 10^{14})$ eV (from radio to multi-TeV range), was fitted semi-analytically by a log-parabola spectrum of $\sim$ TeV electrons \citep{Fraschetti.Pohl:17a,Fraschetti.Pohl:17b}.

The aim of this paper is to formulate a model for particle acceleration and escape in the presence of self-generated turbulence close to the shock and pre-existing turbulence far upstream in order to establish a theoretical foundation of empirical models for energetic particle spectra. A self-consistent theoretical model based on the transport equation coupling the energetic ions and the intensity of upstream ions-generated waves for quasi-parallel magnetic obliquity was built by \cite{Lee:83} and phenomenologically modified to fill the pitch-angle resonance gap by \cite{Ng.Reames:94}. \cite{Kennel.etal:86} found evidence that a correlation between energetic ions and the upstream wave energy density may exist at interplanetary shocks. \cite{Trattner.etal:94} found such a correlation for 300 energetic ion events at the Earth bow shock. Even for local quasi-perpendicular magnetic obliquity, the upstream medium has been long found to deviate from a laminar structure, both for the Earth bow shock \citep{Fairfield:74} and for  interplanetary shocks, as shown by {\it Wind} spacecraft {\it in-situ} measurements of large amplitude whistlers precursors \citep{WilsonIII.etal:17}.

In this paper we present the solution to a new 1-D steady-state transport equation that includes a catastrophic loss term due to particle escape in the presence of self-generated turbulence that matches the pre-existing far upstream turbulence. 
Steady-state 1-D solutions with a catastrophic loss term incorporating ionization and Coulomb losses, nuclear collisions and adiabatic deceleration behind an expanding shock and with a uniform stationary source were presented early-on by \cite{Voelk.etal:81}; a catastrophic loss term describing particle escape from finite extent CME-driven shocks was used to derive broken-power-law momentum spectra by \cite{Li.etal:05}. 

This paper is organized as follows. In Sect. \ref{sec:FEB_escape} the relation between the free escape boundary and the energy dependent escape time is illustrated. In Sect. \ref{sec:outline_model} the transport model used herein is outlined with assumed spatial and energy-dependence of the spatial diffusion coefficient. In Sect. \ref{sec:Intensity_profile} 
we determine analytically the effect on the intensity profile of the particle escape at any energy, not only the highest energy, both from upstream and downstream of the shock for distinct escape times at the shock and distinct particle momenta. In Sect. \ref{sec:spectrum}, we determine the effect of the escape on the steady-state 1-D momentum spectrum calculated at the shock, within an ion-inertial length, and compare it with the log-parabola and the broken power law models. In Sect. \ref{sec:discussion} the results are discussed and Sect. \ref{sec:conclusion} concludes the paper. The Appendix shows that a power law + exponential cut off can be retrieved within the transport model presented herein by assuming a diffusion coefficient uniform throughout except the discontinuity across the shock.

\section{Free escape boundary and escape time} \label{sec:FEB_escape}

In the literature, the upstream particle escape has been customarily modelled by phenomenologically introducing an energy-independent spatial boundary (free-escape boundary, hereafter FEB). Beyond the FEB scattering at any particle energy does not efficiently confine particles that can therefore decouple from the shock and escape with no return to it. Such a spatial cutoff is introduced as in the steady-state infinitely planar version of DSA, accelerated particles, once escaped, are ultimately caught up by the shock that is proceeding at constant speed, whereas particles continue to scatter off the far upstream turbulence. Realistic models should include finite shock lifetime, shock deceleration, geometrical effects, large scale spherical shape or small scale corrugation of the shock surface. 

The FEB was investigated and numerically implemented by several authors via Monte-Carlo simulations \citep{Jones.Ellison:91,Vladimirov.etal:06}, synthesized upstream turbulence test-particle simulations \citep{Giacalone:05a}, 1D spherical simulations with cosmic-rays self-generated turbulence \citep{Kang.Jones:06}, MHD simulations \citep{Reville.etal:08} with instabilities driven by non-resonant streaming cosmic-rays \citep{Bell:04}; overall, the FEB describes the shock as a leaking finite-size system, beside allowing for a considerable reduction of computational time. However, the location of the FEB, i.e.,  $x_{FEB}$, is implemented as independent of the particle momentum. As a result, the relative volume density of low- to high-momentum particles, i.e., the momentum spectrum, at every location in steady-state is artificially affected by the choice of $x_{FEB}$, since higher momentum particles would diffuse out to larger distances than low energy particles due to the larger average square displacement, i.e., diffusion coefficient, that allows them to diffuse further away from the shock with respect to lower energy particles and still be able to return to it; such a return is prevented by an energy-independent FEB. \cite{Drury:11} pointed out the dichotomy resulting from introducing an artificial spatial cutoff such as the FEB: models should merge into a unified picture two competing effects, namely  the decrease of acceleration efficiency, due for instance to the shock deceleration, and gradual particle escape, occurring at all energies, not only the highest ones. The aim of this paper is to outline a viable merge of the two processes.  

A second consequence of the energy-independence of $x_{FEB}$ can be seen as follows. The location $x_{FEB}$, in addition to providing the upstream region of influence of the shock, can be related to a diffusive escape time that can be expressed as $\Delta t (p) \propto x_{FEB}^2/\kappa(p)$ where $\kappa(p)$ is the momentum-dependent spatial diffusion coefficient. 
In the modelling of {\it in-situ} intensity profiles, a momentum-independent $x_{FEB}$ is in tension with the  measurements, that show that even far upstream the intensity profile at distinct particles energy flattens down to the  background level at distinct distances from the shock \citep[e.g.,][]{Lario.etal:19}. Thus, a momentum-dependence of $x_{FEB}$, or in other terms a momentum-dependence of $\Delta t (p)$ distinct from the momentum-dependence of $\kappa^{-1}$, ensures a more realistic description of the intensity profiles that allows particles at distinct energy to escape at distinct locations, not at a unique $x_{FEB}$. Such an effect ultimately changes also the upstream steady-state spectrum. We aim at relating spectral signatures of particle escape (softening) with shock properties (e.g., compression, speed) both in single spacecraft spectra and in multi-spacecraft time-integrated fluences \citep{Mewaldt.etal:12}. 

From a shock with infinite lifetime and constant speed no particle can disappear as all particles diffusing upstream are eventually caught up; however, neither assumption applies to real shocks. Here we introduce an escape time to account for the two forementioned limitations of the DSA. The particle escape is mimicked here by introducing the escape time scale $T(x, p)$ that depends on both position and momentum. We note that in the transport equation the two separate terms of spatial diffusion and escape (in the form of catastrophic losses) are not redundant as each one carries distinct physical information: upstream of a constant speed shock the only diffusion coefficient, despite dependent on momentum and increasing with the distance from the shock, does not entail that, during the shock lifetime, escaped particles will ever return to the shock. Some of them are scattered back and might return to it and others escape. The herein introduced $T(x, p)$ allows to differentiate those that scatter back and return to the shock, keeping a diffusive scattering at the shock, from those genuinely, i.e., observationally, escaped. 

In the downstream plasma, the diffusion competes with both escape and flow advection: at each location $x$, for $T$ longer than the advection time scale, i.e., $x/U$ (where $x$ is the average direction of the shock motion and $U$ the shock speed in the downstream plasma frame), the advection dominates and particles are advected with the flow although some can backscatter and return to the shock (for such a last sub-population, $T$ would be the relevant time scale rather than $x/U$ so that those particles are accounted for with a proper choice of $T>x/U$). The opposite case, namely $T<x/U$, describes high energy particles with velocity component along the flow much greater than $U$.  As shown in the Sect. \ref{sec:Intensity_profile}, the resulting spatial profile differs considerably from the uniform profile predicted by DSA, due to a drop steeper for shorter $T$. In summary, in the 1D case developed below the downstream escape incorporated in $T$ can account both for particles so much faster than the bulk flow speed that diffusion can not efficiently confine them and for backscattering particles able to return to the shock. Geometrical effects such as finite extension or corrugation of the shock surface are not incorporated in $T$ here and require a 2D approach. 

\section{Outline of the model}\label{sec:outline_model}

We consider an infinitely planar shock wave. We use the simplifying assumption that the number of particles per unit time undergoing injection and acceleration processes is balanced by the number of particles per unit time escaping the system so that the steady-state condition holds: $\partial f / \partial t =0$, where $f(x, p)$ is the phase-space distribution function for the energetic particles. Realistic models should describe also the time-dependent imbalance between acceleration and escape; for the sake of simplicity, this effect is neglected herein. We can cast the 1-D steady-state transport equation, assuming pitch-angle isotropy in the local plasma frame, as:

\begin{equation}
U  \frac{\partial f (x,p)}{\partial x} =
\frac{\partial}{\partial x}
\left[ \kappa(x,p)  \frac{\partial}{\partial x} f(x,p) \right] +
\frac{1}{3} \left(\frac{d U}{d x}\right)
~p~\frac{\partial f(x,p)}{\partial p} + S(x,p) - \frac{f (x,p)}{T(x, p)} \, ,
\label{eq:trans}
\end{equation}
where $S(x,p)$ is the source term and the flow speed in the shock frame is uniform both upstream and downstream and discontinuous at the shock on the energetic ions inertial scale:
\begin{equation}
U = \left\{
  \begin{array}{ll}
    U_1   & \mathrm{if~} x <0 , \mathrm{upstream} \, \\
    U_2   & \mathrm{if~} x >0 , \mathrm{downstream}\, .\nonumber
   \end{array}
\right.
\end{equation}

\begin{figure}
	\includegraphics[width=0.9\textwidth]{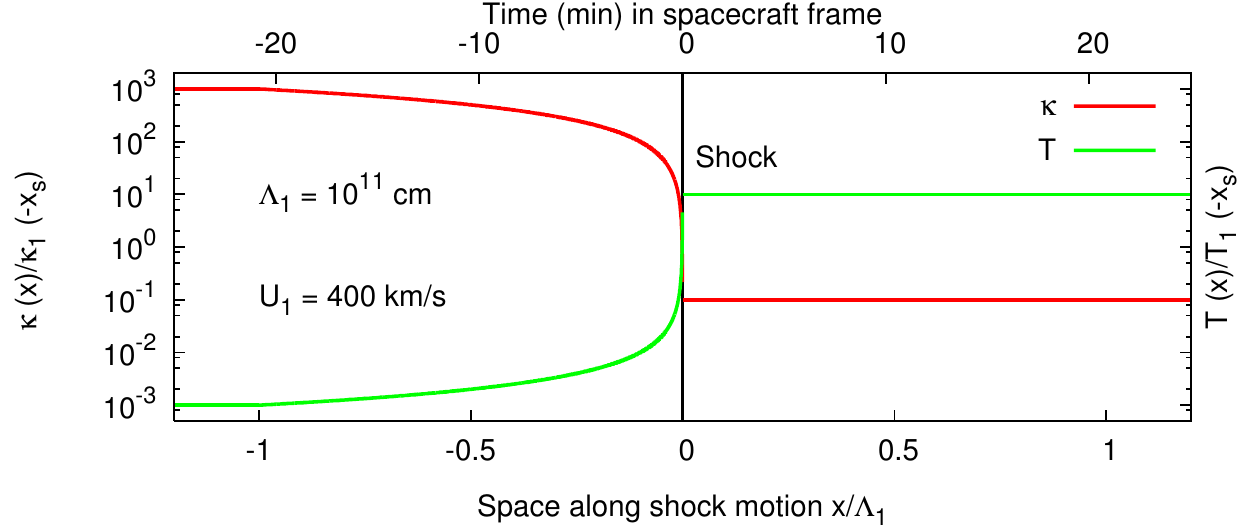}
\caption{Spatial profiles of $\kappa$ and $T$ at a given momentum in units of respective values upstream at distance $|-x_s| = 10\, d_i$ from the shock, where $d_i \simeq 10^7$ cm is the ion inertial length. The upper $x$-axis indicates the time in the spacecraft frame using $U_1 = 400$ km/s at a SW speed $400$ km/s.}
\label{fig0}
\end{figure}

During the last four decades, a vast literature explored the non-linear mechanisms \citep[e.g.,][]{Malkov.Drury:01} exciting the turbulence upstream of shock waves, e.g. resonant  instability by the energetic ions diffusing upstream.  
The density of the scattering centers decreases linearly and $\kappa$ increases linearly upstream for an Alfv\'enic self-generated turbulence \citep{Bell:78a}; for instance, the power spectrum of Alf\'enic fluctuations for high Mach number (supernova remnant) shocks was shown numerically to decrease for each wavenumber over a range of $\sim 4$ decades as the upstream distance from the shock increases \citep{Brose.etal:16}. Far upstream, the pre-existing  interplanetary/interstellar medium  turbulence \citep{Armstrong.etal:95} provides a cutoff out at a certain distance from the shock, i.e., $x = \Lambda_1$, where the self-generated turbulence becomes negligible and $\kappa$ reaches a uniform value, rather than increasing indefinitely (see Fig. \ref{fig0}). The value of $\Lambda_1 = 10^{11}$ cm is consistent with the extension of the suprathermal ($1 -30$ keV) proton foreshocks measured, e.g., in {\it STEREO}  (Solar TErrestrial RElations Observatory) interplanetary shocks \citep[between $0.02$ and $0.1$ AU,][]{Kajdic.etal:12}.

The instability driven by the upstream current of non-resonant energetic ions \citep{Bell:04} generates distinct, i.e., non-linear and non-Alfv\'enic, fluctuations that contributes to reducing the $\kappa$ of the lower-energy ions. The power spectrum of such fluctuations for high Mach number shocks was shown by Monte Carlo simulations to decrease for each wavenumber over a range of $\sim 8$ decades as the upstream distance from the shock increases \citep{Bykov.etal:14}. However, in the absence of a simple functional dependence of $\kappa$ on position inferred from {\it in-situ} measurements or simulations, we assume a linear increase of the upstream $\kappa$ \citep{Bell:78a}. 

The downstream turbulence is advected with the fluid with a typical length-scale much greater than the upstream diffusion scale, with a corresponding weak spatial dependence of $\kappa$; thus, $\kappa$ is assumed to be uniform here. Upstream large-scale density inhomogeneities lead to exponentially rapid amplification of the downstream magnetic field  \citep{Giacalone.Jokipii:07,Fraschetti:13,Fraschetti:14}; for fluctuations length scale as large as the gyroscale of $\lesssim$ GeV protons at $1$ AU (namely $0.07$ AU), such a generated turbulence  might introduce a spatial dependence of the downstream $\kappa$. However, the field amplification scales with the Alfv\'en Mach number $M_A$ \citep{Fraschetti:13} and is modest at the $M_A<10$ interplanetary  shocks\footnote{In contrast, at supernova remnant shocks the amplified turbulent field downstream is stronger ($M_A \simeq 100-1,000$), the gyro-scale of the highest energy protons ($\sim 10^{14} - 10^{15}$ eV) is comparable with the scale of the inhomogeneities and the downstream $\kappa$ might be spatially dependent.}. In addition, the specific location of the most efficient particle acceleration process (whether at the shock or downstream) is still topic of debate. Transport models have interpreted the measured rise of energetic particle intensity downstream of interplanetary shocks \citep{Lario.etal:03} as footprint of an additional source of acceleration therein due to magnetic reconnection \citep{Zank.etal:14,Zank.etal:15}. Also,  measurements of {\it Voyagers} in the  heliosheath, between the termination shock and the heliopause, show a rise of anomalous cosmic rays ($\sim 10-100$ keV particles) with a peak at a distance of $\sim 1$ AU from the shock \citep{Zhao.etal:19}, hence questioning the relative role of the shock transition layer in the acceleration. Equation \ref{eq:trans} does not include a downstream source of acceleration.

We simplify the problem by assuming that $\kappa$ is separable in the following hybrid dependence (see Fig. \ref{fig0}):
\begin{equation}
\kappa(x,p) = \left\{
  \begin{array}{ll}
     \kappa_1(p) \frac{|x - \epsilon|}{|\Lambda_1|}   & \mathrm{if~} x < 0~ {\rm and}~ |x| < |\Lambda_1|, {\rm upstream} \, \\
     \kappa_1(p)    & \mathrm{if~} x < 0~ {\rm and}~ |x| > |\Lambda_1|, {\rm far \,  upstream} \, \\
     \kappa_2(p)    & \mathrm{if~} x > 0 , {\rm downstream} \, ,
    \label{kappa_spec}
   \end{array}
\right.
\end{equation}
where $\kappa_i(p)$ depends only on momentum and $|x - \epsilon|/|\Lambda_1|$ only on space, $i=1,2$ indicates respectively upstream and downstream, $\epsilon = d_i \simeq 10^7$ cm is the ion inertial length. Here $\kappa \rightarrow \kappa_1 (p)$ for $x \rightarrow \Lambda_1$ is the far upstream diffusion coefficient in the pre-existing turbulence. We assume also the separability of $T$; the  upstream $T$ is expected to decrease with the distance from the shock due to the smaller density of scattering centers far from the shock, and higher close to it, whereas the  downstream $T$ is expected to be uniform, mimicking the $\kappa$ dependence. Thus, 
\begin{equation}
T (x,p) = \left\{
  \begin{array}{ll}
	T_1(p)\frac{|\Lambda_1|}{|x - \epsilon|}   & \mathrm{if~} x < 0~ {\rm and}~ |x| < |\Lambda_1|, {\rm  upstream} \, \\
     	T_1(p)   & \mathrm{if~} x < 0~ {\rm and}~ |x| > |\Lambda_1|, {\rm far \,  upstream} \, \\ 
	T_2(p)   & \mathrm{if~} x > 0 , {\rm downstream}. 
    \label{Tesc_spec}
   \end{array}
\right.
\end{equation}

The equations in this Sect. can be compared with the FEB approach that assumes that $f(x_{FEB})=0$ and that the diffusive flux is described by $\kappa(x) \, \partial f / \partial x |_{x=x_{FEB}}$. The solution presented below uses the fact that, from Eqs. \ref{kappa_spec}, \ref{Tesc_spec}, the product $\kappa_i \,T_i$ is independent of space, although the upstream value $\kappa_1 \,T_1$ might differ from the downstream value $\kappa_2 \,T_2$.

\section{Energetic particles intensity profile} \label{sec:Intensity_profile}

\subsection{Upstream intensity profile }

The transport Eq. \ref{eq:trans} with the assumptions outlined in Sect. \ref{sec:outline_model} can be solved analytically in position space. The general solution upstream ($U=U_1$ so that $dU/dx =0$ and far from the source $S(x, p)=0$) is found by solving 
\begin{equation}
U_1  \frac{\partial f (x,p)}{\partial x} =
\frac{\partial}{\partial x} \left[ \kappa_1(x, p) 
 \frac{\partial}{\partial x} f(x,p) \right]  - \frac{f (x,p)}{T(x, p)} .
\label{eq_kappa_var_up}
\end{equation}
We factorize $f$ as $f(x)=u(x) \, v(x)$, where $u(x)$ and $v(x)$ are solutions respectively to the equations
\begin{equation}
\frac{\partial^2 u}{\partial x^2} + \left[ -\frac{1}{\kappa_1(x, p) T_1(x, p)} - \frac{4 \nu_1^2 -1}{4 x^2} \right] u = 0  , \qquad
\frac{\partial^2 v}{\partial x^2} + {\frac{1}{2}} \left[ \frac{\partial a}{\partial x} - \frac{a^2}{2} \right] v = 0  , \nonumber
\end{equation}
where  $\nu_1 (p) = |\Lambda_1|U_1/2 \kappa_1(p)$ and $a = (\partial \kappa_1/\partial x  - U_1)/\kappa_1$. The solution to the $u$-equation is a linear combination of Bessel functions \citep[see][Eq. 8.491]{Gradshteyn.Ryzhik:07}. The phase-space distribution function $f(x,p)$ in the upstream can be recast as
\begin{equation}
f(x, p) =  \underbrace{\left[C_1 J_\nu \left(\frac{i |x-\epsilon|}{ L_1 (p) }\right) + C_2 Y_\nu \left(\frac{i |x-\epsilon|}{L_1 (p)}\right) \right] }_\text{u(x)} \, \underbrace{ {\rm exp} \left(-\int_x^0 \frac{U (x')}{2 \kappa (x', p)} d x' \right)}_\text{v(x)}
\end{equation}
where $L_1 (p) \equiv \sqrt{\kappa_1(x, p)T_1(x, p)} = \sqrt{\kappa_1(p)T_1(p)}$ is defined here as the escape length, independent of the distance from the shock, $J_\nu$, $Y_\nu$ are Bessel functions of first kind and second kind, respectively, $i$ is the imaginary unit and $C_1$, $C_2$ are constants to be determined by boundary conditions; the last exponential factor is reminiscent of the upstream steady-state 1D solution for an infinitely planar shock in the case of no escape (DSA). Since the spatial dependencies of $\kappa$ and $T$ are expected to be reciprocal or {\it nearly} reciprocal, a choice of such dependencies different from Eqs. \ref{kappa_spec} and  \ref{Tesc_spec} might introduce at most a weak spatial dependence of $L_1 (p)$. 

By imposing that $f(x, p)$ is not exponentially divergent far upstream ($f(x,p) \not\rightarrow \infty$ as $x \rightarrow -\infty$) leads to the relation $C_2 = i C_1$; this is found by using the asymptotic expressions for $J_\nu$, $Y_\nu$ \citep[see][Eq.s 8.451.1 - 8.451.2]{Gradshteyn.Ryzhik:07} and taking only the term $k=0$ of the $k$-expansion. 

The complex-valued constant $C_1$ is determined by imposing $f (x,p)=f_0(p)$ at the shock ($x=0$). We use the series representations of $J_\nu$, $Y_\nu$ \citep[see][Eq.s 8.440 - 8.443 - 8.451.2]{Gradshteyn.Ryzhik:07} and take only the term $k=0$ of the expansion since $x \simeq 0$. 

The exact upstream solution to Eq. \ref{eq:trans}, where the integration constants are calculated using the boundary conditions and approximations described above, can then be recast as
\begin{equation}
f(x, p) = f_0 (p)\, \Gamma(1 - \nu_1) \, \left( \frac{|x-\epsilon|}{2L_1 (p)} \right)^{\nu_1} {\cal R}  \left[ - {\rm e}^{-i \pi \nu_1} I_{\nu_1}\left( -\frac{|x-\epsilon|}{L_1 (p)} \right) + I_{-\nu_1}\left(\frac{|x-\epsilon|}{L_1 (p)} \right) \right]
\label{sol_up}
 \end{equation}
where $I_{\nu_1} (z)$ is the modified Bessel functions of imaginary argument \citep[see][Eq. 8.406.1]{Gradshteyn.Ryzhik:07}, $\Gamma (z)$ is the gamma function and ${\cal R} [.]$ indicates the real part of $[.]$.
The relevant limits of Eq. \ref{sol_up} are analysed below.

The asymptotic limit of Eq. \ref{sol_up} for $|x-\epsilon| \gg L_1(p)$, by using the asymptotic expression of $I_{\nu}(z)$ in \cite{Gradshteyn.Ryzhik:07}, Eq. 8.451.5, and taking only the term $k=0$, leads to the following upstream profile
\begin{equation}
f(x, p) \rightarrow f_0 (p) \, \frac{\Gamma(1 - \nu_1)}{\sqrt{\pi}} \, {\rm sin} ( \pi \nu_1) \, \left( \frac{|x-\epsilon|}{2L_1 (p)} \right)^{\nu_1 - 1/2}  {\rm exp}\left( -\frac{|x-\epsilon|}{L_1 (p)} \right) \, , {\rm upstream} .
\label{sol_up_infty}
\end{equation}
We note that the $\Lambda_1$-dependence of the upstream profile $f(x, p)$ is implicit in $\nu_1(p)$. The spatial profile in Eq. \ref{sol_up_infty} has to be compared with the steady state 1D test-particle solution of DSA with no upstream escape. We readily note that for large distance from the shock, i.e., limit $|x| \gg L_1$ (and $|x| < |\Lambda_1|$), $f$ is exponentially suppressed with a roll-over scale $L_1 (p)$, that replaces the roll-over scale in the case of no escape (DSA), i.e., $\sim \kappa_1(\epsilon, p)/U_1$. Thus, particles do not disappear and still spread out to an infinite distance as in DSA, but with an exponentially small amplitude.

The DSA upstream profile is recovered if, in the surrounding of a certain far upstream location $|\bar x| \gg |x_s|$, the escape time becomes comparable with the acceleration time: 
\begin{equation}
T_1(\bar x, p) \simeq \frac{\kappa_1(\bar x, p) }{U_1^2} \, ,
\label{bal_esc_acc}
\end{equation}
where the right-hand side of Eq. \ref{bal_esc_acc} is a crude estimate of the acceleration time \citep{Axford:81,Forman.Drury:83,Drury:83}. This limit can be clarified as follows. If the condition in Eq. \ref{bal_esc_acc}, equivalent to $\kappa_1 (\bar x, p)/U_1 \simeq L_1(p)$, is satisfied, the last  exponential factor of Eq. \ref{sol_up_infty} becomes ${\rm exp} \,[-U_1 |\bar x - \epsilon|/\kappa_1(\bar x, p)]$, i.e., the profile in the case of no-escape: ${\rm exp}(-\int_x ^0 dx' \,U_1/\kappa_1(x', p) \simeq -U_1 \Delta x/\kappa_1(\bar x, p)$, where $\kappa_1(\bar x, p)$ is interpreted as an average of $\kappa_1$ within the upstream interval $(x, 0)$. The only additional spatially-dependent factor in Eq. \ref{sol_up_infty}, i.e., $(|x-\epsilon|/2L_1 (p) )^{\nu_1 - 1/2}$, absent in the no-escape solution, tends to zero for large $|x|$ as $\nu_1 < 1/2$ for the particle energy of interest herein:
\begin{equation}
\nu_1 (p_0) = 0.1 \frac{|\Lambda_1|}{10^{11}\, {\rm cm}} \frac{U_1}{200\, {\rm km/s}} \left(\frac{\bar \kappa_1}{10^{19} \, {\rm cm^2/s}} \right)^{-1} \, ;
\label{eq:nu1}
\end{equation}
where we have introduced the normalization constant $\bar \kappa_1 = \kappa_1 (\Lambda_1, p_0)$; the typical value of $\kappa_1(\Lambda_1, p_0) \simeq 10^{19} \, {\rm cm^2/s}$ for $p_0$ corresponding to $\sim 100$ keV protons is consistent with the mean free path measured across the interplanetary medium \citep{Palmer:82}. Equation \ref{eq:nu1} shows that even for very fast shocks and small particle momentum (small $\kappa$), the inequality $\nu_1 < 1/2$ is likely to be satisfied. 
We note that the assumptions herein lead to a ratio of acceleration-to-escape time scale,  $\kappa_1(x, p)/U_1^2 T_1(x, p)$, increasing quadratically with the distance from the shock, as  $\kappa_1 \propto x$ and $T_1 \propto 1/x$ (see Eqs. \ref{kappa_spec}, \ref{Tesc_spec}); thus, the likelihood of upstream escape is continuously varying as particles progress upstream rather than being governed by an artificial switch operating only at $x = x_{FEB}$. 

Illustrative monochromatic ($p=p_0$) intensity profiles are shown in Fig. \ref{fig1}, upper panel, for $T_1 (-x_s) = 400$ sec upstream at a distance from the shock $|-x_s| = 10 \, d_i$ and distinct ratios $T_2/T_1(-x_s)$. The diffusion coefficient for $\sim 100$ keV protons at the shocks is chosen as $\kappa_1 (-x_s, p_0) = 10^{16}$ cm$^2/$s; smaller values of $\kappa_1$ \citep[leading to higher acceleration rates, ][]{Jokipii:82,Jokipii:87} are not ruled out at local high obliquity shocks. The DSA profile is drawn for comparison (in cyan) with an upstream  $\kappa_1 (\bar x, p) = \kappa_1^{DSA} = 10^{17}$ cm$^2/$s so that the diffusion scale is $\kappa_1^{DSA} /U_1$ equals the escape length $L_1(p_0) = 2 \times 10^9$ cm. The value of $\kappa_1^{DSA}$ is consistent with the recent determination by \cite{Kis.etal:18} of the e-folding distance ahead of quasi-parallel obliquity regions of the Earth bow shock for $\sim 10-40$ keV protons. The insert in the upper panel in Fig. \ref{fig1} shows that the upstream escape depletes the particle intensity very close to the shock, within $L_1(p_0)$, to $\sim 20 \% $ of the DSA prediction due to the factor $(|x-\epsilon|/2L_1 (p) )^{\nu_1 - 1/2}$ (see Eq. \ref{sol_up_infty}); far upstream the intensity dilutes asymptotically with comparable slope, in logarithmic scale, as DSA due to $\kappa_1^{DSA} /U_1 = L_1(p_0)$. An analysis \citep{Kis.etal:04} of  multi-spacecraft $10-32$ keV ion events at the Earth bow shock led to an estimate of the roll-over upstream distance shorter than the statistical expectation \citep{Trattner.etal:94}. The profile determined in Eq. \ref{sol_up_infty} suggests an efficient escape as explanation of the short roll-over distance, in alternative to the explanation based on a large SW velocity \citep{Kis.etal:04}.  

The lower panel in Fig. \ref{fig1} shows that a long escape time ($T_1 (-x_s) = 8,000$ sec, blue line) leads  to a rapid  drop within the DSA diffusion length ($\kappa_1^{DSA}/U_1$, corresponding to $\sim 0.37$ min in spacecraft frame); the shallower decrease of the blue curve far upstream, that leads to the crossing of the blue curve with the DSA curve, results from the increase of $\kappa_1$ with distance from the shock that increases the effective diffusion length far from the shock.  
Equation \ref{bal_esc_acc}, equivalent to $\kappa_1 (\bar x, p)/U_1 \simeq L_1(p)$, shows that if $T_1(p)$ is smaller than the acceleration time scale at $x=\bar x$, $L_1$ is the length scale of the roll-over profile; if $T_1(p)$ becomes comparable to the acceleration time scale, the escape is not a dominant process and the profile is governed by acceleration and diffusion, as in the case of DSA. We note that if $T$ and $\kappa$ have different momentum dependence, the threshold in Eq. \ref{bal_esc_acc} is met at distinct locations for distinct momenta. Thus, Eq. \ref{bal_esc_acc} illustrates quantitatively the balance between escape and acceleration.

We note that the condition of no-divergence far upstream imposed on $f(x, p)$ is consistent with an escape time scale shorter than the acceleration time scale at large distance from the shock, thereby efficient escape from the shock over a broad range of distances from the shock for distinct particle energy. 
From the assumed dependencies in Eq. \ref{kappa_spec}, it holds (for $|-x_s| < |\Lambda_1|$)
\begin{equation}
{\rm exp} \left(-\int_x^0 \frac{U (x')}{2 \kappa (x', p)} d x' \right)  = \left( \frac{|x-\epsilon|}{\epsilon} \right)^{\nu_1} .
\nonumber
\end{equation}
By using the asymptotic expressions for the Bessel functions \citep[see][Eq.s 8.451.1 - 8.451.2]{Gradshteyn.Ryzhik:07}, we can recast the spatially dependent factors of $f(x, p)$ in Eq. \ref{sol_up_infty} as 
\begin{equation}
\left( \frac{|x-\epsilon|}{\epsilon} \right)^{\nu_1}{\rm exp} \left(- \frac{|x-\epsilon|}{L_1 (p)} \right) =   {\rm exp}\left( \nu_1 \, {\rm ln} \frac{|x-\epsilon|}{\epsilon} -\frac{|x-\epsilon|}{L_1 (p)} \right) .
\label{exp_fact_2}
\end{equation}
Although the left hand side in Eq. \ref{exp_fact_2} is clearly exponentially suppressed for any value of the parameters $\nu_1$, $\kappa_1 (p)$ and $T_1 (p)$, it is useful to note that the condition
\begin{equation}
 \nu_1 \, {\rm ln} \frac{|x-\epsilon|}{\epsilon} -\frac{|x-\epsilon|}{L_1 (p)} \lesssim 0 \, ,
\nonumber
\end{equation}
that is necessary to prevent exponential divergence, implies far upstream 
\begin{equation}
T_1(p) \lesssim  \frac{\kappa_1(p)}{U_1^2} \, .
\label{eq:times_acc_esc}
\end{equation}
Equation \ref{eq:times_acc_esc} links the mathematical condition of no-divergence of the solution with the fact that acceleration cannot balance escape and particles at any momentum are allowed to escape, a different times $T_1 (p)$ (cfr. Eq. \ref{bal_esc_acc}), again emphasizing that the approach presented herein describes the continuous process of escape at any particle energy. 

We note that, although $L_1$ does not depend explicitly on the advection speed $U_1$ unlike the diffusion scale, the escape in this 1D case depends on the balance between advection and diffusive confinement (confinement is assumed to be infinitely efficient at any particle energy in the case of no escape within a given FEB), as in the DSA. In the presence of 2D shock the geometry is expected to contribute the particles  confinement. 

Finally, from a mathematical viewpoint, at large distance from the shock $f$ has to tend to zero and cannot be a finite constant since the only uniform solution of Eq. \ref{eq:trans} is the identically zero distribution. This is at odds with the case of no escape (i.e., $f(x, p)/T \sim 0$ in Eq. \ref{eq:trans}): a non-vanishing constant depending only on momentum is the asymptotic upstream solution. 

\subsection{Downstream intensity profile }\label{sec:down_int}
  
In the downstream region ($U=U_2$ so that $dU/dx =0$ and far from the source $S(x, p)=0$), we  
impose continuity with the upstream solution, i.e., $f(0, p) = f_0(p)$; then,  
taking the limit for $x \rightarrow + \infty$ and discarding the exponentially divergent solution yields
\begin{equation}
f(x,p) = f_0 (p) \, {\rm exp} \left[ \left( \frac{U_2}{2\kappa_2(p)} - {\frac{1}{2}} \sqrt{\left(\frac{U_2}{\kappa_2(p)}\right)^2 + \frac{4}{L_2^2 (p)}} \right) x \right] \, , {\rm downstream},
\label{spat_down_unif}
\end{equation}
where $L_2 (p) \equiv  \sqrt{\kappa_2(p)T_2(p)}$ is the downstream escape-length. In the limit of no-escape ($T_2(p) \gg \kappa_2(p)/U_2^2$ or, equivalently, $L_2 (p) \gg \kappa_2(p)/U_2$ ), the downstream profile tends to the uniform limit $f_0 (p)$, recovering the DSA solution. Equations \ref{sol_up_infty} and \ref{spat_down_unif} show that the profiles depend on the ratio $ \kappa (p)/U L(p)$ (upstream coinciding with the ratio of the diffusion length to the escape length): if $T (p)$ is very large the DSA solution is retrieved.

\begin{figure}
	\includegraphics[width=0.8\textwidth]{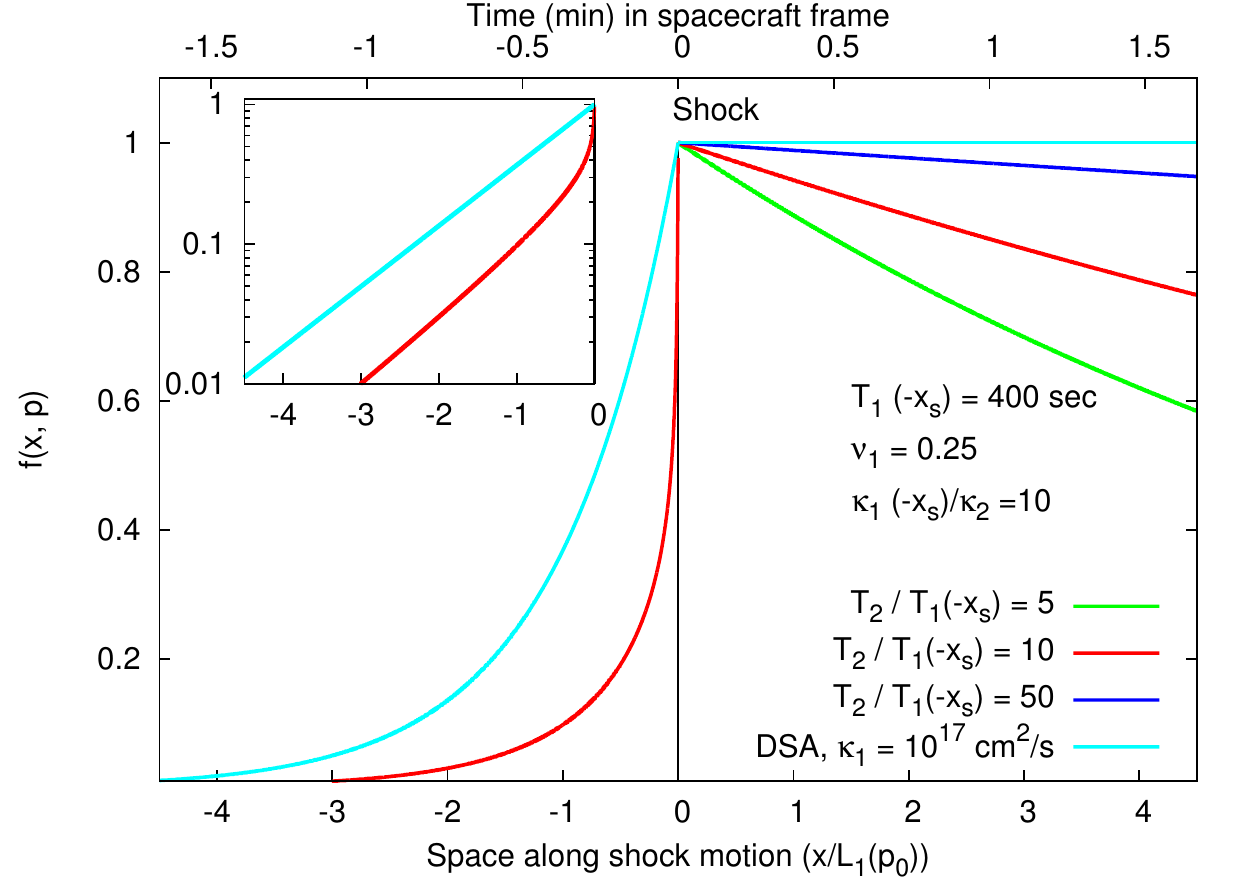}
	\includegraphics[width=0.8\textwidth]{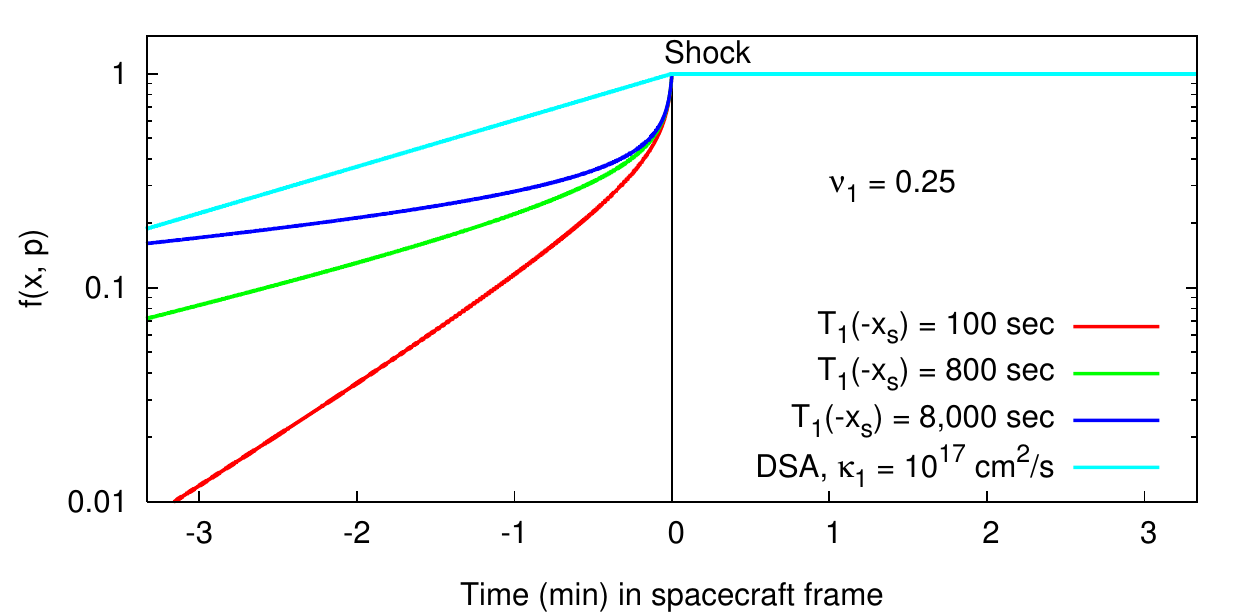}
\caption{{\it Upper panel:} Intensity profiles of energetic particles upstream (Eq. \ref{sol_up_infty}) and downstream (Eq. \ref{spat_down_unif}) of the shock, normalized to the shock value, for distinct ratios $T_2 (p_0)/T_1(-x_s, p_0)$ at a given momentum $p_0$ and fixed $\kappa_1 (-x_s, p_0)/\kappa_2 (p_0) =10$. The DSA profile is shown (in cyan) for comparison with $\kappa_1^{DSA} = 10^{17}$ cm$^2/$s. For $\sim$100 keV protons, we have used $T_1 (-x_s, p_0) = 400$ sec, $\kappa_1 (-x_s, p_0) = 10^{16}$ cm$^2/$s and $U_1 = 500$ km$/$s so that $\kappa_1^{DSA} /U_1 = L_1(p_0) = 2 \times 10^9$ cm. The red curves correspond to the case $\kappa_1 (p_0)/\kappa_2 (p_0) = T_1 (p_0)/T_2 (p_0) = 10$ so that $L_1 (p_0)= L_2 (p_0)$. The shock compression is $r=3$. The vertical black line marks the shock location. The lower $x$-axis gives the distance along the shock normal in units of $L_1(p_0)$; the upper $x$-axis gives the time in spacecraft frame (assuming a SW speed $= 400$ km/s). The panel inserted in the top left zooms into the upstream region in logarithmic scale on the $y$-axis. {\it Lower panel:} Same as upper panel comparing upstream profiles only (Eq. \ref{sol_up_infty}) with DSA profile for distinct $T_1(-x_s, p_0)$. The $x$-axis gives the time in spacecraft frame.}
\label{fig1}
\end{figure}

In the downstream region, Fig. \ref{fig1} (upper panel) shows that large escape times ($T_2 > 2,000$ sec) as compared to the advection time ($\sim$ a few minutes) leads the profile to the DSA flat shape, as the escape becomes irrelevant and the profile is advection-dominated. As $T_2$ is shortened and becomes comparable to the advection time, the profile drops more and more steeply behind the shock due to particles moving downstream away from the shock and faster than the advected flow.

Figure \ref{fig1}, upper panel, shows intensity profiles for fixed $\kappa_1 (-x_s, p_0)/\kappa_2 (p_0) =10$ and distinct values of $T_2 (p_0)/T_1 (-x_s, p_0)$ within the shock layer. The case $\kappa_1 (-x_s, p_0)/\kappa_2 (p_0) = T_2 (p_0)/T_1 (-x_s, p_0) = 10$, i.e., $L_1 (p_0)= L_2 (p_0)$, is shown by the upstream and downstream red curves. For this case the upstream curve falls steeper than the DSA curve (cyan curve in the insert) close to the shock but reaches the same asymptotic slope; the downstream profile is steeply declining, in contrast with the DSA case. Only increasing $T_2/T_1$ at fixed $\kappa_1/\kappa_2$ ($L_1<L_2$) leads to a flatter downstream profile closer to the DSA prediction. Thus, a comparison with DSA only, suggests $L_1<L_2$; however, measured upstream profiles steeper than DSA suggest that the case $L_1>L_2$ can be realised. High time-resolution spacecraft data at terrestrial bow shock or interplanetary shocks can help investigate and constrain this model. Figure \ref{fig1}, upper panel, shows roll-over timescale of $\sim 1$-minute; the high cadence of, e.g. the {\it Magnetospheric Multiscale} ({\it MMS}) mission, {\it Hot Plasma} and {\it Energetic Particles Detector} \citep{Cohen.etal:19}, allows an accurate estimate of the shock parameters and of energetic particles profile within such a time interval.

\subsection{Profile for distinct particle  momentum}\label{sec:p_int}

\begin{figure}
		\includegraphics[width=0.7\textwidth]{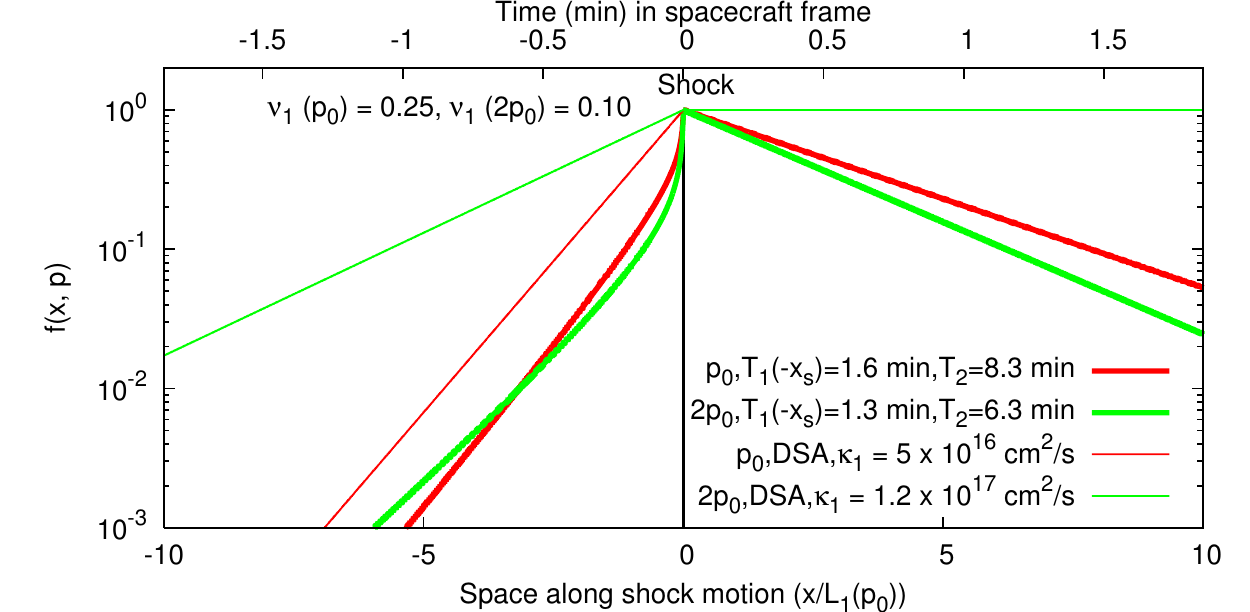}
\caption{Intensity profiles for distinct particle momenta $p_0$ (corresponding to $\sim$100 keV, in thick red) and $2\, p_0$ (in thick  green). Here  $\kappa_1 (-x_s, p_0) = 10^{16}$ cm$^2/$s, $\delta_1 =1.3$, $ T_1 (-x_s, p_0) = 100$ sec, $\gamma_1 = 0.4$ and $U_1 = 500$ km$/$s; we also use $\delta_1=\delta_2$,$\gamma_1=\gamma_2$,  $\kappa_1 (-x_s, p_0)/ \kappa_2 (p_0) = 10$ and $T_2(p_0)/T_1(-x_s, p_0) = 10$. For comparison, the DSA profiles for $p=p_0, 2\, p_0$ are shown as thin curves in red and green, respectively, with scaling $\kappa_1^{DSA}(p/p_0)^{\delta_1}$ and such that $\kappa_1^{DSA} (p_0) /U_1 = L_1 (p_0)= 10^9$ cm. The shock compression is $r=3$. The vertical black line marks the shock location. The lower $x$-axis gives the distance along the shock normal in units of $L_1(p_0)$; the upper $x$-axis gives the time in spacecraft frame (assuming a SW speed $= 400$ km/s). }
\label{fig2}
\end{figure}

Figure \ref{fig2} shows the dependence of the intensity profile on $p$ under the assumption of a power-law $p$-dependence of $\kappa$ and $T$:
\begin{equation}
\kappa_i (p) = \bar \kappa_i (p/p_0)^{\delta_i} , \quad T_i (p) = \bar T_i (p/p_0)^{-\gamma_i} , \quad \delta_1 > \gamma_1
\label{kappa_T_p}
\end{equation}
where $\bar \kappa_1$ is defined below Eq. \ref{eq:nu1}, $\bar \kappa_2 = \kappa (p_0)$, $\bar T_1 = T_1 (\Lambda_1, p_0)$, $\bar T_2 = T_2 (p_0)$ and power-law indexes $\delta_i >0$, $\gamma_i >0$; the condition $\delta_1 > \gamma_1$ is motivated below. The value of $\delta_i$ depends on the particular turbulence assumed in the upstream medium ($\delta_i = 4/3$ in the 3D isotropic turbulence and $\delta_i = 1$ in the Bohm diffusion limit). The value of $\gamma_i$ is relatively unconstrained. In this figure we keep for the sake of simplicity  $\delta_1 = \delta_2$ and $\gamma_1 = \gamma_2$, neglecting the effects of compression or magnetic field amplification at the shock. For two distinct values of momentum ($p_0$ and $2 \, p_0$) and $T_1 (-x_s, p_0) = 100$ sec, Fig. \ref{fig2} compares the intensity profiles with the respective DSA prediction such that $\kappa_1^{DSA} (p_0) /U_1 = L_1 (p_0)= 10^9$ cm, assuming the same $\delta_1$. Larger momenta lead to larger diffusion length ($\kappa_1^{DSA}/U_1$); likewise, the condition $\delta > \gamma$ leads to an increasing $L_1(p) \propto (p/p_0)^{(\delta_1-\gamma_1)/2}$ with momentum (as in Fig. \ref{fig2}), thereby explaining the crossing of the thick red and green curves upstream: the factor $\left( \frac{|x-\epsilon|}{2L_1 (p)} \right)^{\nu_1 - 1/2}$ in Eq. \ref{sol_up_infty} steepens the profile close to the shock at larger $p$ due to the $\nu_1(p)$ dependence (see Eq.\ref{eq:nu1}) but $L_1(p)$ is also larger for larger $p$ thereby diluting the profile far upstream. The profiles in Fig. \ref{fig2} are generally in agreement with the spacecraft measurements that higher energy particles diffuse out to larger distances from the shock than lower energy particles. Thus, the case $\delta < \gamma$ (leading to $L_1(p)$ also smaller for larger $p$) is ruled out by {\it in-situ} measurements. In addition, we note that the assumptions in Eqs. \ref{kappa_T_p} also imply that the ratio $\kappa_1(x, p)/U_1^2 T_1(x, p)$ increases with a power-law index in momentum $\delta_1 + \gamma_1 > 1$ (see Eq.\ref{kappa_T_p}), satisfied for any value $\gamma$ for the turbulence models mentioned above ($\delta \geq 1$), consistently with the expectation that for higher $p$ the escape is favoured over the acceleration.

\section{Momentum spectrum}\label{sec:spectrum}
 
From the continuity of $f(x, p)$ across the  shock, the momentum spectrum at the shock $f_0 (p) = f(p)$ can be derived following the usual textbook procedure in the case of DSA. The conservation of the number of particles flowing along the $x$-direction across the shock, i.e., $\int_{-\epsilon} ^{+\epsilon} dx \, U \partial f / \partial x = 0$, applied to Eq. \ref{eq:trans} reads:
\begin{equation}
\left[ \kappa(p)  \frac{\partial f(x,p)}{\partial x}   +
\frac{1}{3} U p~\frac{\partial f(x,p)}{\partial p}\right] \bigg\lvert_{-\epsilon} ^{+\epsilon} + \int_{-\epsilon} ^{+\epsilon} dx \left[ S(x,p) - \frac{f (x,p)}{T(p)} \right]= 0 .
\label{contin_unif}
\end{equation}

From the upstream exact solution for a spatially varying $\kappa_1 (x,p)$ in Eq. \ref{sol_up} and the recursion formula for the derivative of $I_\nu (z)$ \citep[see Eq. 8.486.1]{Gradshteyn.Ryzhik:07}, it holds
\begin{equation}
\frac{\partial f (x, p)}{ \partial x}  \bigg\lvert_{-\epsilon} \simeq f_0 (p) \frac{\Gamma(1 - \nu_1(p))}{\Gamma(1 + \nu_1(p))} \frac{\nu_1 (p)}{2L_1 (p)}  \, \left( \frac{\epsilon}{2L_1 (p)} \right)^{2\nu_1(p) - 1}  
\label{gradf_var_up}
\end{equation}
where we have used $\epsilon/L_1 (p) \ll 1$ and the $p$-dependence of $\nu_1$ through $\kappa_1$ is made explicit. 

As for the downstream region, the gradient of $f(x, p)$ from Eqs \ref{spat_down_unif}:
\begin{equation}
\frac{\partial f(x,p)}{\partial x} \bigg\lvert_{+\epsilon} =  \left( \frac{U_2}{2\kappa_2 (p)} - {1\over 2} \sqrt{\left(\frac{U_2}{\kappa_2(p)}\right)^2 + \frac{4}{L_2^2 (p)}} \right) f(x, p) \, . 
\label{gradf_unif_down}
\end{equation}

For $\nu_1 <1$, we use the approximation 
\begin{equation}
\frac{\Gamma(1 - \nu_1)}{\Gamma(1 + \nu_1)} = \frac{\pi}{\nu_1 \Gamma^2(\nu_1) {\rm sin} (\pi \nu_1)} \simeq  \frac{1}{\nu_1 ^2 \Gamma^2(\nu_1)} \, . 
\end{equation}
The source is assumed to be monochromatic and localized at the shock: $S(x,p) = S_0 \delta(x) \delta(p-p_0)$. The two gradients in Eqs. \ref{gradf_var_up} and \ref{gradf_unif_down}, replaced into Eq. \ref{contin_unif}, yield the following equation for $f(p)$, in the limit for $\epsilon \rightarrow 0$:
\begin{equation}
\frac{\partial f(p)}{\partial p} - \frac{3r}{p U_1 (1-r)} \left[\frac{\kappa_1 (p)}{|\Lambda_1|} \frac{2}{\nu_1 \Gamma^2(\nu_1(p))} \left(\frac{\epsilon}{2 L_1 (p)} \right)^{2\nu_1(p)}
- \frac{U_2}{2} \left(1  - \sqrt{1+\frac{4 \kappa_2(p)}{U_2^2 T_2(p)}}\right) \right] f(p) + {S_0 \over U_1} \frac{3r}{1-r} \frac{\delta(p-p_0)}{p} =0 \, .
\label{eq:spect_cont}
\end{equation}
By approximating $\Gamma(\nu_1(p)) \sim {1\over \nu_1(p)} - C$, where $C$ is the Euler constant \citep[][Eq. 8.322]{Gradshteyn.Ryzhik:07}, and using $\kappa_2(p)/U_2^2 T_2(p) \ll 1$, we can recast the dominant terms of the solution of Eq. \ref{eq:spect_cont} as
\begin{equation}
f(p) \propto \underbrace{ \left( {p\over p_0} \right)^{-q} \, {\rm exp} \left[ -A_1 \left( {p\over p_0} \right)^{-\delta_1} \right]  }_\text{upstream} 
\underbrace{{\rm exp} \left[ -A_2 \left( {p\over p_0} \right)^{\delta_2+\gamma_2} \right]  }_\text{downstream} \, , 
\label{spec_hybrid_comp_T}
\end{equation}
where the symbol ``$\propto$'' refers to the ${\cal O}(1)$ factors of proportionality with momentum-dependence $(p/p_0)^{(p/p_0)^{-\delta_1}}$, ${\rm exp}(-(p/p_0)^{-\delta_1})$ not included in Eq. \ref{spec_hybrid_comp_T},  the factor $p^{-q}$, with the constant exponent $q = 3r/(r-1)$, is the DSA test-particle power law in the case of no-escape and
\begin{eqnarray}
A_1 &=& 0.4 \, \frac{q}{\delta_1} \frac{U_1}{400 \, {\rm km/s}} \frac{|\Lambda_1|}{10^{11} \, {\rm cm}} \left(\frac{\bar \kappa_1}{10^{19}  \,{\rm cm^2/s}}\right)^{-1} \left|-5.3 + {\rm log} \frac{\epsilon/10^7 {\rm cm}}{ L_1(p_0)/10^{9} {\rm cm}}  \right| , \\   
A_2 &=& 6.25 \times 10^{-5}  \frac{q}{r(\delta_2 + \gamma_2)} \frac{\bar \kappa_2}{10^{15} \,{\rm cm^2/s}} \left(\frac{U_2}{400 \, {\rm km/s}}\right)^{-2} \left(\frac{\bar T_2}{10^4 \,{\rm sec}}\right)^{-1} 
\label{eq:A1_A2}
\end{eqnarray}
where we have used $\bar T_2 = \bar T_1(0, p_0)$. We note that the expression for $f(p)$ in Eq. \ref{spec_hybrid_comp_T} does not depend on $\gamma_1$: the momentum spectrum does not depend on the assumption on the upstream power law index of $T$ in momentum provided that $\delta_1 > \gamma_1$ and $\delta_1 + \gamma_1 > 1$.  The width of the peak of the spectrum at low $p$ is controlled by the constant $A_1$, namely by $U_1 \Lambda_1 ({\rm log} \, L_1)/\bar \kappa_1$, whereas the drop at large $p$ is controlled by the constant $A_2$, namely by $\bar \kappa_2 /U_2^2 \bar T_2$. Table \ref{tab:1} lists some values of $A_1$, $A_2$ for illustrative purpose.

\begin{figure}
	\includegraphics[width=1.1\textwidth]{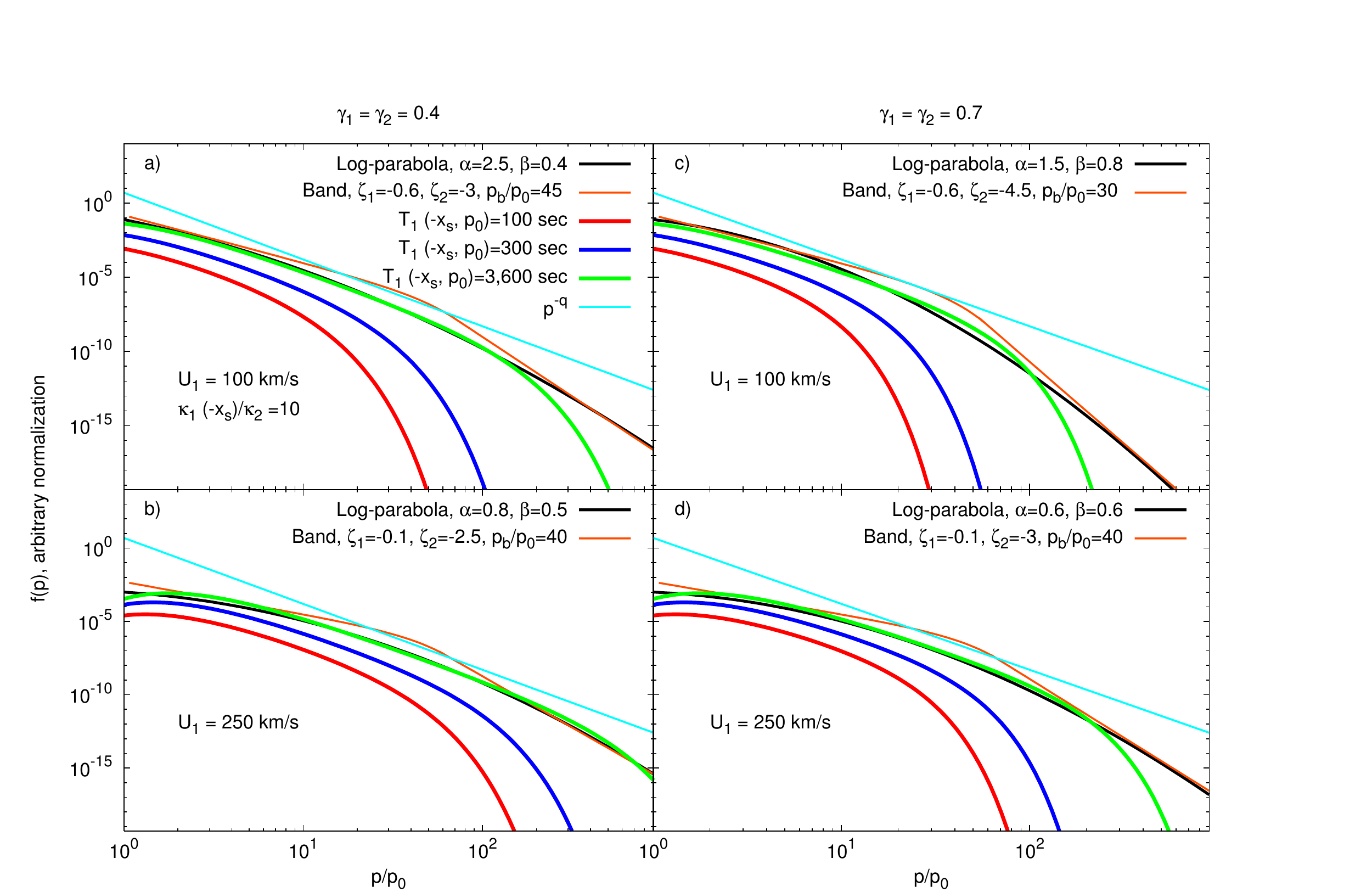}  
\caption{{\it a)}: Momentum spectrum in Eq. \ref{spec_hybrid_comp_T} for distinct values of $T_1 (-x_s,p_0)$ $=100$ sec (in red), $300$ sec (in blue) and $3,600$ sec (in green),} compared with the test-particle DSA solution (in cyan), with a log-parabola (in black) reproducing the spectrum in the case $T_1 (-x_s,p_0) = 3,600$ sec (in this panel up to $p/p_0 \simeq 150$ that corresponds to $\simeq 4$ decades in particle energy), and with a Band function (in orange) for illustrative purpose with asymptotic slopes fitting the log-parabola. Here $\bar \kappa_1=10^{19}$ cm$^2$/s, $|\Lambda_1| = 10^{11}$ cm, $r=3$, $U_1 = 100$ km/s, $\delta_1 = \delta_2 =1$, $\gamma_1= \gamma_2 = 0.4$, $\kappa_1 (-x_s, p_0)/\kappa_2(p_0) = 10$ and in all cases $T_2 (p_0)/T_1(-x_s, p_0) =10$ so that $\kappa_2(p)/U_2^2 T_2(p) \ll 1$. The blue and red curves are multiplied by arbitrary factors, respectively by $0.1$ and $0.01$, for readability purposes. {\it b)}: Same as {\it a)} with $U_1 = 250$ km/s. {\it c)}: Same as {\it a)} with $\gamma_1= \gamma_2 = 0.7$.  {\it d)}: Same as {\it c)} with  $U_1 = 250$ km/s. 
\label{fig4}
\end{figure}

Figure \ref{fig4} compares the spectrum in Eq. \ref{spec_hybrid_comp_T} for distinct $T_1 (-x_s,p_0)$ with DSA test-particle power-law and with the log-parabola spectrum ($\propto (p/p_0)^{-\alpha - \beta\, {\rm log}(p/p_0)}$) that fits in all panels the green curve, corresponding to the case $T_1 (-x_s,p_0) = 3,600$ sec, where the curvatures $\beta$ span the range of the best-fit values for the spectra of the 16 GLE events during solar cycle $24$ \citep{Zhou.etal:18}. Figure \ref{fig4} also depicts the phase-space distribution function $f_B(p)$, corresponding to the Band function customarily used to fit the measured differential intensity, i.e., $dJ/dE = p^2 \, f_B(p)$ for non-relativistic energy, that can be recast as 
\begin{equation}
f_B (p) \propto \left\{
  \begin{array}{ll}
    (p/p_0)^{2\,(\zeta_1-1)} \, {\rm exp} (-(p/p_b)^2)  & \mathrm{if~} p < \sqrt{\zeta_1 -\zeta_2}\, p_b \\
    (p/p_0)^{2\,(\zeta_2-1)} \, [(\zeta_1-\zeta_2) (p_b/p_0)^2]^{\zeta_1 - \zeta_2} \, {\rm exp} (\zeta_2 - \zeta_1)  & \mathrm{if~} p > \sqrt{\zeta_1 -\zeta_2} \, p_b\, , 
   \end{array}
\right.
\end{equation}
where $\zeta_1$ and $\zeta_2$ are power law indexes of $dJ/dE$ as a function of the energy rather than $p$. In Fig. \ref{fig4} $f_B(p)$ is chosen to fit the  asymptotic slopes of the log-parabola at low- and large-momentum .

The parameters of the log-parabola $\alpha, \beta$ are related to the probability of particle escape from the acceleration region; in particular, the curvature $\beta$ is related to $R=E/E_0$, i.e., the energy gain in the single shock-crossing with initial (final) energy $E_0$ ($E$): $\beta = q'/2 \, {\rm log}\, R$, where $q'$ is a parameter of the probability of escape ${\cal P} (E) = g/E^{q'}$ where $g$ is a constant \citep{Massaro.etal:04,Fraschetti.Pohl:17a}. 

The determination of the microscopic parameters $\alpha$, $\beta$ (and $\zeta_1$, $\zeta_2$) in terms of the new constants $A_1$, $A_2$, that depend on macroscopic transport and shock parameters, requires the determination of constant coefficients in a transcendental equation. The graphic solution in Fig. \ref{fig4} suggests the best-fit correspondence between $\alpha, \beta$ (or $\zeta_1$, $\zeta_2$) and $A_1$, $A_2$, $\delta_i$ for distinct $\gamma_i$. Figure \ref{fig4}, b)  shows that the asymptotic slopes of the Band curve emerge, over almost $3$ decades in momentum (i.e., $> 5$ decades in kinetic energy), for large values of $T_1 (-x_s, p) \simeq 1 $ hour, large $U_1 \simeq 250$ km/s, small index $\gamma_2=0.4$ and $\kappa_1 (-x_s, p_0)/\kappa_2(p_0) = 10$. In all panels the Band function overestimates the spectrum from both other models at $p \simeq p_b$, allowing to discriminate between models\footnote{The dependence of the energy break on the ions charge-to-mass ratio \citep{Mewaldt:06,Desai.etal:16} can be expressed as dependence of $\beta$ on the ions charge-to-mass ratio via the anti-correlation $ \beta \simeq E_b ^{-0.36}$, determined only for the 16 GLE of solar cycle 24 \citep{Zhou.etal:18}, where $E_b$ is the Band function energy break.}  \citep{Zhou.etal:18}. Likewise, in the escape-free DSA, the spectral power-law index can be expressed both as a function of the macroscopic density compression and, alternatively, as a function of microscopic quantities, i.e., of the probability of remaining in the acceleration region ($P$) and of $R$: ${\rm log} P/{\rm log} R$.

 \begin{table}[h!]
\begin{center}
\begin{tabular}{ |c|c|c|c|c|c|c|c| } 
\hline
\multicolumn{2}{|c|}{} & \multicolumn{2}{|c|}{$\gamma_1= \gamma_2=0.4$} & \multicolumn{4}{|c|}{\hspace{0.15cm} $\gamma_1= \gamma_2=0.7 $} \\
\hline
$U_1$ (km/s)& $T_1(-x_s,p_0)$ (s) & $A_1$ & $A_2$ &  $A_1$ & $A_2$ \\
\hline
\multirow{3}{4em}{$100$} & $100$ & $2.38$ & 0.0964 &  $2.38$ & $0.0794$ \\ 
& $300$ & $2.63$ & $0.0321$ & $2.63$ & $0.0265$ \\ 
& $3,600$ & $3.19$ & $0.00268$ &  $3.19$ & $0.00221$\\ 
\hline
\multirow{3}{4em}{$250$} & $100$ & $5.96$ & $0.0154$ & $5.96$ & $0.0127$ \\ 
& $300$ & $6.58$ & $0.00514$ & $6.58$ & $0.00424$ \\ 
& $3,600$ & $7.98$ & $0.000429$ &  $7.98$ & $0.000353$\\ 
\hline
\end{tabular}
\caption{Spectral parameters $A_1$, $A_2$ for the cases depicted in Fig. \ref{fig4} are summarized in terms of $U_1$, $T_1$ and $\gamma_i$.}
\label{tab:1}
\end{center}
\end{table}

\section{Discussion}\label{sec:discussion}

The solution in Eq. \ref{spec_hybrid_comp_T} provides a simple no-power-law momentum spectrum of energetic particles, remarkably different from the power-law + exponential cutoff and can be mapped into a log-parabola. The log-parabola parameters $\alpha$ and $\beta$ and the Band function parameters $\zeta_1$, $\zeta_2$ and $p_b$ are related via $A_1$, $A_2$ (see Eq.s \ref{eq:A1_A2}) to the DSA test-particle power-law index $q$, the upstream shock speed $U_1$, the far-upstream spatial diffusion coefficient $\bar \kappa_1$, the length-scale $\Lambda_1$ where the self-generated turbulence is drowned into the pre-existing far upstream turbulence and logarithmically by the escape length $L_1(p_0)$. In contrast, in the power-law + exponential cutoff solution the roll-over length-scale depends on the diffusion properties close to the shock. Due to the weak (logarithmic) dependence of $A_1$ on $L_1(p_0)$, the value of $L_1(p_0)$ is likely to be constrained via the {\it in-situ} measured intensity profiles rather than the spectra. In the Appendix the power-law+exponential cutoff spectrum \citep{Ellison.Ramaty:85} is shown to be retrieved in the limit of acceleration faster than escape ($\kappa_1(p)/U_1^2 T_1(p) \ll 1 $) by solving Eq. \ref{eq:trans} with $\kappa$ and $T$ spatially independent throughout, provided a discontinuous jump at the shock. 

The value of $\gamma$ is  constrained by the two requirements that for higher $p$ (see Sect. \ref{sec:p_int}) the intensity profiles extend further upstream ($\gamma < \delta$, or $dL/dp > 0$) and that the ratio of acceleration to escape time scale grows with momentum ($\delta_1 + \gamma_1 > 1$, realized for a 3D isotropic turbulence or in the Bohm limit). Further constraints on $T$ and on its spatial dependence both at the shock and far from it can be identified via direct numerical simulations that are beyond the scope of this work. In addition, $\gamma$ is likely to be affected by the nature of the magnetic turbulence, as much as  $\delta$; a specific turbulence power spectrum is not implemented herein for the sake of simplicity. The fact that $\delta \neq \gamma$ might seem at odds with the assumption that the upstream spatial dependences of $\kappa$ and $T$ are reciprocal: if both $\kappa$ and $T$ are governed by the same turbulence, $\delta$ and $\gamma$ should be nearly equal. An argument supporting the reciprocal spatial dependence, beside the higher density of scattering centers close to the shock, is that the only quantity with dimension of length that combines $\kappa$ and $T$, that is $L(p)$, is expected to be weakly dependent on space (in Eqs. \ref{kappa_spec} and \ref{Tesc_spec} it is exactly independent of $x$), as much as the diffusion length (for a uniform $U_1$); on the other hand, $L_1(p) \propto (p/p_0)^{(\delta_1-\gamma_1)/2}$ is expected to depend on $p$ to account for the observed larger extent of the diffusive streaming ahead of the shock at larger $p$; thus, $\delta_1 \neq \gamma_1$. Alternatively, a momentum-dependence of the escape time can be derived from confinement arguments, by assuming a Sedov-type time dependence for the maximum momentum of the energetic particles before leaving the shock, with no return to it both from upstream and from downstream  \citep{Ptuskin.Zirakashvili:05,Celli.etal:19b}. 

In the context of the escape of cosmic rays from the galaxy, \cite{Lerche.Schlickeiser:85} concluded that an agreement with the observed trend of the secondaries-to-primaries cosmic rays flux ratio at energies $>10$ GeV/nucleon can be achieved, for primaries accelerated both by shocks and by momentum diffusion off Alfv\'enic turbulence, only if both $\delta$ and $\gamma$ (respectively $\eta$ and $b$ therein) are non-zero and positive. The case $\delta \neq \gamma$ for the primary cosmic rays was not ruled out by observations. Modelling of non-thermal electron emission in solar flares also makes use of a distinct  momentum-scaling of $\kappa$ and $T$ \citep{Petrosian:16}.

The model introduced herein can be regarded as an extension of the leaky-box model. Typically an escape time is introduced to solve separately the transport equation in the momentum space and enters as eigenvalue in a series expansion of the steady-state transport equation with a FEB \citep[chap. 14]{Schlickeiser:02}. In the model presented here, the space and momentum dependence of $\kappa$ prevents the separate solution of the transport equations for position and momentum. Nevertheless, it is shown here that a spatial solution can be determined analytically, if the upstream is populated by a self-generated turbulence that at a distance $|\Lambda_1|$ from the shock is taken over by pre-existing turbulence rather than vanishing at a FEB.

Discontinuous changes in the $f(x, p)$ such as charge exchange (or leptons pair annihilation) are not included in the term  $f/T$ in Eq. \ref{eq:trans}.  For protons, the charge exchange with thermal upstream hydrogen atoms, using a SW density $\sim 1$ cm$^{-3}$ and a relative speed $\lesssim 1,000$  km/s that is much larger than the shock speed in the frame of the upstream interplanetary medium, has\footnote{We use the International Atomic Energy Agency database \texttt{http://www-amdis.iaea.org/ALADDIN/}.} time scale $\sim 10^6$ sec.  For charge exchange protons/heavier ions (such as C,O), such a time scale is of order $10^7 -10^8$ sec. Thus, the charge exchange time scales are  much longer than any time scale of interest for interplanetary shocks; however, they are relevant for longer-lived (e.g., interstellar) shocks. For ions fragmentation due to ion-ion collisions or spallation, the time scale is much longer than $\sim$years at the SW density and in the particle energy range considered here \citep[$0.1  - 1$ GeV;][]{Silberberg.Tsao:90}. These processes are therefore also not included in $T$ for interplanetary shocks although they need to be included for interstellar shocks. 

We propose that the particles escaped from the shock and diffusing into the SW are described by a diffusion equation that does contain the loss term $f/T$ hereby introduced. However, the scale of the acceleration region that particles escape from might exceed by several orders of magnitude the scale of the shock region, close to the scale of the heliosphere, thereby making the $f/T$ term negligible: the escape time is much larger than any other relevant time scale for processes occurring during the  propagation of the particles, e.g., not only charge exchange but also adiabatic energy losses, reacceleration by another shock, trapping into planetary magnetosphere, etc. In addition, the escape term might be relevant to identify the source of the shock-accelerated particles measured as SEP (at $1$ AU but also at {\it Parker Solar Probe} or {\it Solar Orbiter } much closer to the Sun
) arriving hours before the shock that produced them. Equation \ref{eq:trans} can be helpful to model such an ``SEP escape''.

Finally, the approach presented herein does not aim at describing the formation of the spectral energy tail from an initial single Maxwellian distribution; this process is egregiously being tackled by Particle-In-Cell (PIC) simulations despite over limited spatial and temporal domains \citep[see e.g.,][and references therein]{Pohl.etal:20}. On the other hand the gyroradius and energy scales of the majority of the escaping particles are beyond reach of the current PIC capabilities. Test-particles numerical simulations are a suitable investigation tool to measure numerically the absolute value and the scaling in position and momentum of the $T$ as herein introduced, but they are beyond the scope of this work.

\section{Conclusion}\label{sec:conclusion}

We have built a steady-state 1D transport model for energetic particles at shocks allowing for a finite  acceleration-to-escape time scale ratio upstream at any distance from the shock and at any particle energy, with no imposed energy-independent escape boundary. By using a hybrid spatial dependence of the diffusion coefficient, i.e., linearly increasing with the distance from the shock upstream to account for the self-generated turbulence due to streaming ions \citep{Bell:78a} and uniform downstream, we find intensity profiles and momentum spectra of energetic particles remarkably different from the DSA solution. As for the intensity profiles, the steep drop upstream of the shock suggests an interpretation of the supra-thermal ions events in the Earth bow shock  multi-spacecraft observations \citep{Kis.etal:04}. Although a large SW local velocity might explain the short roll-over distance in front of the shock, the model for the escape presented here offers a possible alternative that deserves to be explored and compared with data in detail. Far upstream, at a distance comparable with the diffusion scale, the profile recovers the DSA exponential roll-over, although with a reduced amplitude. The downstream profile is uniform (as in DSA) for downstream escape time larger than the advection time and drops from the shock more steeply than DSA as the escape time becomes comparable with the local advection time.

We have provided a derivation of the 1D momentum spectrum that has the form of the product of a power law and  two exponentials and can be mapped into a log-parabola. This model offers a derivation of the log-parabola spectrum complementary to the probability-based derivation used, e.g., in \cite{Massaro.etal:04,Fraschetti.Pohl:17a}. In summary, the power-law scaling of the momentum spectrum, that results from a process of particle energization increasing at higher energy \citep[dating back to the original idea in][]{Fermi:49} is extended herein to a more general scaling of a power law modulated by two exponentials; such a new form of the spectrum matches within a certain parameter range the log-parabola spectrum, that results from allowing particle escape at any energy/distance from the shock, not only at the highest energy nor only at a given spatial boundary. We emphasize that the log-parabola does not replace the power law, applicable in narrower energy ranges, but rather extends it to a broader energy range.

Multi-dimensional effects (shock rippling, large-scale shock curvature, finite extent of the shock surface not considered here) also contribute to reshaping the high-energy part of the spectrum and were considered in  \cite{Drury:11,Malkov.Aharonian:19}. The effect of shock corrugation at the highest energy ion scale on the escape can be analytically included in this model \citep{Fraschetti:13,Fraschetti:14} and comparison with multi-spacecraft  measurements can be used to constrain the momentum dependence of $T$. 

The prospect of measuring with {\it Parker Solar Probe} and {\it Solar Orbiter} the escape of $0.1 - 1$ GeV particles åccelerated at very high speed shocks, more likely to be crossed close to the Sun than at $1$ AU, opens new opportunities to test the model herein presented. So far only a handful of remarkably weak shock events have crossed either spacecraft, but the ongoing increase of the solar activity toward the next maximum is expected to lead to observations in an unchartered physical domain.

\acknowledgments

The constructive comments of the referee are gratefully acknowledged. This work was supported, in part, by NSF under grant 1850774, by NASA under Grant 80NSSC18K1213 and by {\it Chandra} Theory grant GO8-19015X. FF thanks Drs. J. Giacalone and J. Kota for discussions and suggestions and Drs. A. M. Bykov, L. Drury and M. W. Pohl for comments in an email exchange.

\appendix

\section{Spatial profile: case of spatially uniform $\kappa$ and $T$}\label{sec:spatial_uniform}

The shock is assumed not to significantly perturb the medium it is propagating into so that the upstream parameters of the turbulence can be assumed to hold pre-existing values and are unchanged with the distance from the shock. Thus, we can use a spatial diffusion coefficient $\kappa$ uniform in space both upstream and downstream and discontinuous at the shock only: 
\begin{equation}
\kappa (x,p) = \left\{
  \begin{array}{ll}
     \kappa_1(p)   & \mathrm{if}~x < 0 \, , \mathrm{upstream} \, \\
     \kappa_2(p)   & \mathrm{if}~x > 0 \, , \mathrm{downstream} \, .
    \label{kappa_unif}
   \end{array}
\right.
\end{equation}
Likewise, the escape time $T$ is assumed to be discontinuous at the shock and uniform elsewhere: 
\begin{equation}
T (x,p) = \left\{
  \begin{array}{ll}
     T_1(p)   & \mathrm{if}~x < 0 \, , \mathrm{upstream} \\
     T_2(p)   & \mathrm{if}~x > 0 \, , \mathrm{downstream} \, .
    \label{Tesc_unif}
   \end{array}
\right.
\end{equation}
The transport equation \ref{eq:trans} with the assumptions in Eqs. \ref{kappa_unif}, \ref{Tesc_unif} is  solved in coordinates-space as follows. 

\subsection{Upstream spatial profile }

We seek the general solution upstream ($U=U_1$ so that $dU/dx =0$ and far from the source $S(x, p)=0$) of the equation

\begin{equation}
U_1  \frac{\partial f (x,p)}{\partial x} =
\kappa_1(p) 
\left[ \frac{\partial^2}{\partial x^2} f(x,p) \right]  - \frac{f (x,p)}{T(p)} .
\label{eq_kappa_unif_up}
\end{equation}
We impose two boundary conditions to determine the two integration constants. The first one gives the value of $f$ at the shock $f(0, p) = f_0(p)$. 
The second integration constant is chosen to prevent exponential divergence.   
The upstream solution is
\begin{equation}
f(x,p) = f_0(p) \, {\rm exp} \left[ \left( \frac{U_1}{2\kappa_1 (p)} + {1\over 2} \sqrt{\left(\frac{U_1}{\kappa_1(p)}\right)^2 + \frac{4}{L_1^2 (p)}} \right) x \right] 
\label{spat_up_unif}
\end{equation}
that tends to  ${\rm exp} (U_1 \, x/\kappa_1(p)) $ as $T_1(p) \rightarrow \infty$, i.e., with identical spatial dependence as the upstream profile in case of no particle-escape (DSA) and in case of  space-independent $\kappa$ and $U$. Equation \ref{spat_up_unif} shows that if the upstream $\kappa$ is uniform, the effective diffusion length $\left( \frac{U_1}{2\kappa_1 (p)} + {1\over 2} \sqrt{\left(\frac{U_1}{\kappa_1(p)}\right)^2 + \frac{4}{L_1^2 (p)}} \right)^{-1}$ is smaller  than $\kappa_1/U_1$ for any $T_1$. Unlike the case of no-escape, where  the asymptotic limit at $x \rightarrow -\infty$ is a non-vanishing constant depending only on momentum (that also solves Eq. \ref{eq_kappa_unif_up} in the limit $T_1(p) \rightarrow \infty$), the solution in Eq. \ref{spat_up_unif} tends exponentially to zero far upstream, due to the escape. An exponential dilution of particles is also found in the case of spatially dependent $\kappa$ (see Sect. \ref{sec:Intensity_profile}). Likewise, particles do not disappear but spread out to an infinite distance  with an exponentially small amplitude.

\subsection{Downstream spatial profile }
  
In the downstream region  ($U=U_2$ so that $dU/dx =0$ and far from the source $S(x, p)=0$) 
we impose the continuity at the shock with the upstream solution: $f(0, p) = f_0(p)$. 
The second integration constant is again chosen to prevent exponential divergence. The solution is then
\begin{equation}
f(x,p) = f_0 (p) \, {\rm exp} \left[ \left( \frac{U_2}{2\kappa_2(p)} - {1\over 2} \sqrt{\left(\frac{U_2}{\kappa_2(p)}\right)^2 + \frac{4}{L_2^2 (p)}} \right) x \right] 
\label{spat_down_unif_appendix}
\end{equation}
with limit  $f_0 (p)$ in the case of no-escape ($T_2(p) \rightarrow \infty$), as expected. Both upstream and downstream profiles depend on the ratio of $\kappa (p)/U$ (upstream coinciding with the diffusion length) to  $L(p) = \sqrt{\kappa(p) T (p)}$: if  $T (p)$ is very large, one recovers the DSA solution.

\begin{figure}
	\includegraphics[width=0.8\textwidth]{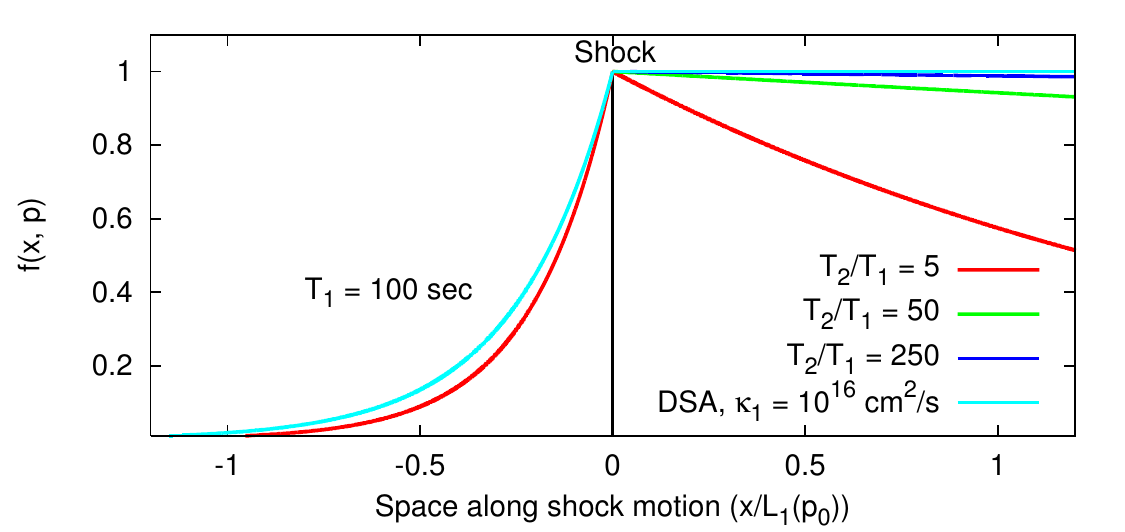}
	\includegraphics[width=0.8\textwidth]{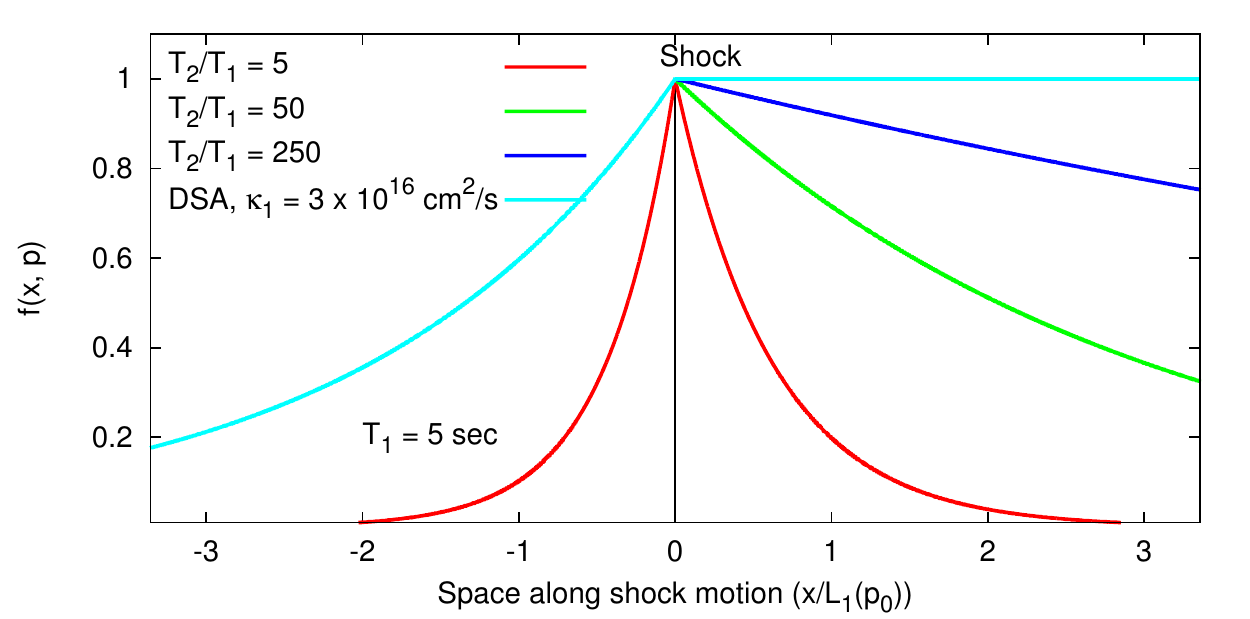}
\caption{{\it Upper panel:} Spatial profiles in the case of uniform $\kappa$ and $T$ for a given momentum $p_0$ of the  accelerated protons (corresponding to $\sim$100 keV) for fixed $T_1 (p_0) = 100$ sec and distinct $T_2/T_1 (p_0)$; here we choose $\kappa_1 (p_0)/\kappa_2 (p_0) =5$, $\kappa_1 (p_0) = 10^{16}$ cm$^2/$s and $U_1 = 400$ km$/$s so that the diffusion scale is $\kappa_1 (p_0) /U_1 = 2.5 \times 10^8$ cm and $L_1 = 10^9 $ cm (see Eq. \ref{spat_up_unif}). The red curves correspond to the case $\kappa_1 (p_0)/\kappa_2 (p_0) = T_1 (p_0)/T_2 (p_0) = 5$ so that $L_1 (p_0)= L_2 (p_0)$. The shock compression is $r=3$. The DSA profile at $p=p_0$ is shown in cyan for comparison. {\it Lower panel:} Same as upper panel with $T_1 (p_0) = 5$ sec, $\kappa_1 (p_0) = 3. \times 10^{16}$ cm$^2/$s and $\kappa_1 (p_0) /U_1 = 7.5 \times 10^8$ cm $> L_1 =3.9 \times 10^8$ cm. }
\label{fig5}
\end{figure}

Examples of spatial profiles are illustrated in Fig. \ref{fig5} for $T_1 = 100$ sec and distinct $T_2$ (upper panel), for $T_1 = 5$ sec and distinct $T_2$ (lower panel), both at a given particle momentum $p_0$ corresponding to $\sim 100$ keV protons; here $\kappa_1 (p_0) = 10^{16}$ cm$^2/$s \citep{Kis.etal:18}. The DSA profile is drawn for comparison (in cyan). The upstream profile is close to the DSA exponential roll-over for large  $T_1 = 100$ sec $\gg x/U_1 \sim 10$ sec (Fig. \ref{fig5}, upper panel): in this case the diffusion scale is $\kappa_1 (p_0) /U_1 = 2.5 \times 10^8$ cm and $L_1 = 10^9 $ cm. For $T_1 =5 \, {\rm sec} < x/U_1 $ (Fig. \ref{fig5}, lower panel) the profile steepens with respect to the DSA prediction (here $L_1 =\sqrt{\kappa_1 T_1} < \kappa_1/U_1$). In both panels, large downstream escape times ($T_2 \gtrsim 1,000$ sec, upper panel,  blue and black curves, and lower panel, black curve) with respect to the advection time $x/U_2 \simeq 30$ sec lead to a profile comparable with the uniform prediction of DSA, as the escape becomes irrelevant and the profile is advection-dominated. For $T_2 \sim 25$ sec $< x/U_2$ (lower panel, green curve) the profile drops behind the shock due to particles moving downstream away from the shock faster than the advection and not efficiently back-scattering to return to the shock.

\section{Momentum spectrum: case of spatially uniform $\kappa$ and $T$}\label{sec:spec_unif}
 
From the continuity of the spatial profiles across the shock (see Sect. \ref{sec:spatial_uniform}), the momentum spectrum at the shock $f_0 (p) = f(p)$ can be found by following the usual textbook derivation in the case of no-escape. The number of particles flowing along the x-direction has to be continuous across the shock: $\int_{-\epsilon} ^{+\epsilon} dx \, U \partial f / \partial x = 0$, for an infinitesimally small $\epsilon$. From Eq. \ref{eq:trans}, the continuity simply reads:
\begin{equation}
\left[ \kappa(p)  \frac{\partial f(x,p)}{\partial x}   +
\frac{1}{3} U p~\frac{\partial f(x,p)}{\partial p}\right] \bigg\lvert_{-\epsilon} ^{+\epsilon} + \int_{-\epsilon} ^{+\epsilon} dx \left[ S(x,p) - \frac{f (x,p)}{T(p)} \right]= 0 .
\label{contin_unif_appendix}
\end{equation}

\begin{figure}
	\includegraphics[width=0.75\textwidth]{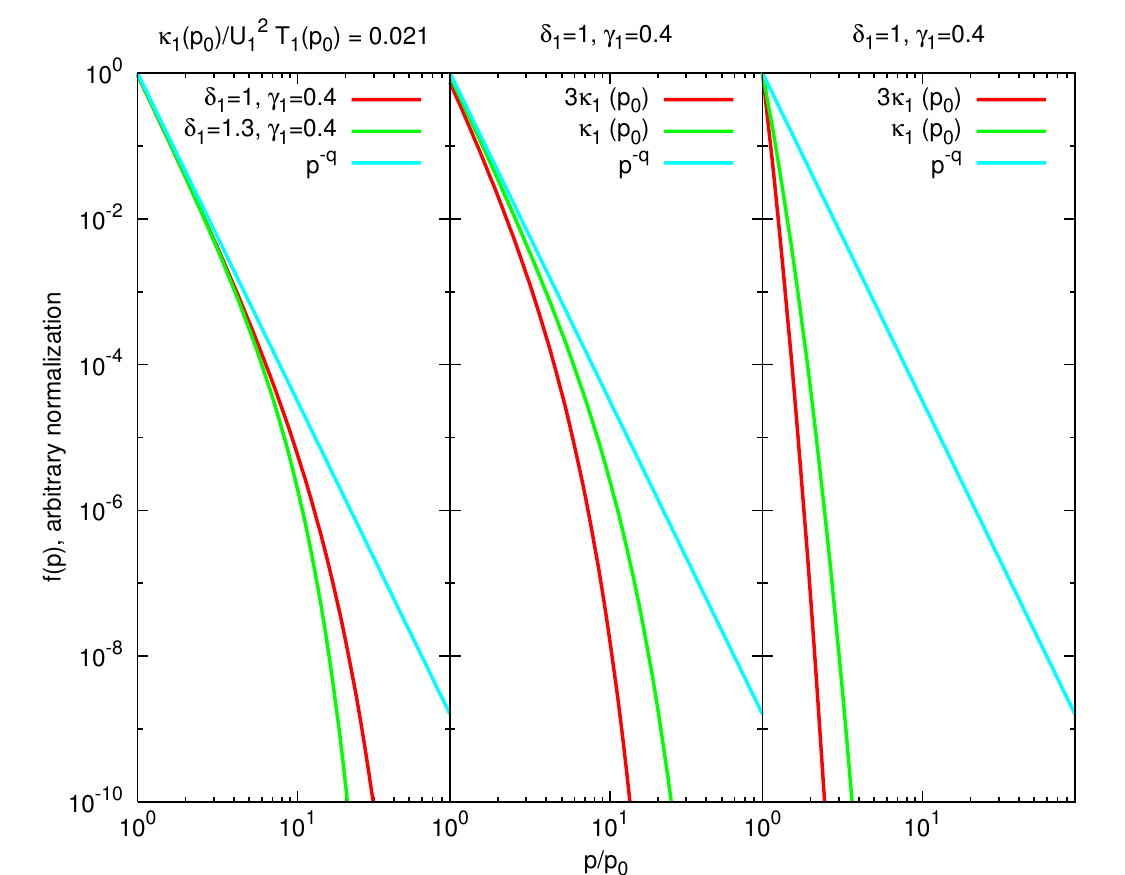}
\caption{{\it Left panel:} Spectrum in the case of uniform $\kappa$ and $T$ for distinct values of $\delta_1$  and $\kappa_1(p_0)/U_1^2 T_1(p_0) \ll 1$, compared with the test-particle DSA solution (in cyan). Here $\kappa_1 (p_0) = 10^{16}$ cm$^2/$s, $T_1 (p_0) = 300$ sec, $\gamma_1=0.4$, $U_1 = 400$ km$/$s and $r=3$.  {\it Middle panel:} Same as left panel for $\kappa_1 (p_0)$, $3 \kappa_1 (p_0)$ and fixed $\delta_1$, $\gamma_1$ , $T_1 (p_0) = 200$ sec, compared with the DSA solution. Other parameters are unchanged from the left panel.  {\it Right panel:} Spectrum in the case of uniform $\kappa$ and $T$ (with $\kappa_1(p_0)/U_1^2 T_1(p_0) \gg 1$) for $\kappa_1 (p_0)$, $3 \kappa_1 (p_0)$. Here $T_1 (p_0) = 20$ sec, $U_1 = 100$ km$/$s, $\delta_1 =1$, $\gamma_1=0.4$. The red and green spectra are normalized to $1$ at $p=p_0$. }
\label{fig6}
\end{figure}

By calculating the upstream and downstream gradients of $f$ from Eqs. \ref{spat_up_unif} and \ref{spat_down_unif_appendix},
using a mono-chromatic source at the shock, i.e., $S(x,p) = S_0 \delta(x) \delta(p-p_0)$, and replacing the two gradients into Eq. \ref{contin_unif_appendix} yields
the following equation for $f(p)$, in the limit of $\epsilon \rightarrow 0$:
\begin{equation}
\frac{\partial f(p)}{\partial p} - \frac{3r}{2p (1-r)} \left[1+\sqrt{1+\frac{4 \kappa_1(p)}{U_1^2 T_1(p)}} - {1\over r}\left( 1+\sqrt{1+\frac{4 \kappa_2(p)}{U_2^2 T_2(p)}}\right) \right] f(p) + {S_0 \over U_1} \frac{3r}{1-r} \frac{\delta(p-p_0)}{p} =0
\end{equation}
where we have used that  $\int_{-\epsilon} ^{+\epsilon} dx  f (x,p)/T(p) = 0$. 
The general solution has a cumbersome logarithmic form. For momentum small enough that the limit $\kappa_i(p)/U_i^2 T_i(p) \ll 1$ is satisfied both upstream and downstream ($i=1,2$), i.e., acceleration time scale shorter than escape time scale, the solution has an interesting form. By assuming the momentum dependence for $\kappa_i(p)$ and $ T_i(p)$ in Eqs. \ref{kappa_T_p},
we recast the solution as
\begin{equation}
f(p) \propto \left( {p\over p_0} \right)^{-q} {\rm exp} \left[ - \frac{q}{\delta_1+ \gamma_1} \frac{\bar \kappa_1}{U_1^2 \bar T_1} \left( {p\over p_0} \right)^{\delta_1 + \gamma_1}  \right] .
\label{spectrum_unif_1}
\end{equation}
where we have used $\kappa_1(p)/T_1(p) \gg \kappa_2(p)/T_2(p)$; the exponential roll-over is governed by power-laws exponents of $\kappa$ and $T$ combined: $\delta_1 + \gamma_1$. In Eq. \ref{spectrum_unif_1} the power-law + exponential roll-over solution \citep{Ellison.Ramaty:85} is retrieved, with steeper drop for larger $\bar \kappa_1$, as expected. 
Figure \ref{fig6} (left and middle panels) depicts the spectrum in Eq. \ref{spectrum_unif_1} for two cases: 1) for fixed $\kappa_1(p_0)$, $\gamma_1$ and distinct $\delta_1$; 2) fixed $\delta_1$ and distinct $\kappa_1(p_0)$.

In the opposite limit, i.e., $\kappa_i(p)/U_i^2 T_i(p) \gg 1$, the resulting spectrum is (see Fig. \ref{fig6}, right panel):
\begin{equation}
f(p) \propto (p/p_0)^{-3/2} {\rm exp} \left[ - \frac{2q}{\delta_1+ \gamma_1} \sqrt{\frac{\bar \kappa_1}{U_1^2 \bar T_1}} (p/p_0)^{\frac{\delta_1 + \gamma_1}{2}}  \right] .
\end{equation}
In this case, the acceleration is so inefficient that the spectrum is suppressed exponentially more vigorously than the solution in Eq. \ref{spectrum_unif_1} (this condition reads $\delta_1 + \gamma_1 < 2$).

\addcontentsline{toc}{section}{Bibliography}

\def \apss{{\it Astrophys.\ Sp.\ Sci.}}
\def \aj{{\it AJ}}
\def \apj{{\it ApJ}}
\def \apjl{{\it ApJL}}
\def \apjs{{\it ApJS}}
\def \araa{{\it Ann. Rev. A \& A}}
\def \prc{{\it Phys.\ Rev.\ C}}
\def \aap{{\it A\&A}}
\def \aaps{{\it A\&ASS}}
\def \grl{{\it Geophys. Res. Lett.}}
\def \jgr{{\it J. Geophys. Res. (Space Physics)}}
\def \jcap{{\it J. Cosmology and Astroparticle Phys.}}
\def \mnras{{\it MNRAS}}
\def \physrep{{\it Phys.\ Rep.}}
\def \physscr{{\it Phys.\ Scripta}}
\def \pasp{{\it Publ.\ Astron.\ Soc.\ Pac.}}
\def \gca{{\it Geochim. Cosmochim.\ Act.}}
\def \nat{{\it Nature}}
\def \solphys{{\it Sol.\ Phys.}}


\bibliographystyle{aasjournal}
\bibliography{ffraschetti}

\end{document}